\documentclass[12pt,english,prb,preprint,superscriptaddress,showpacs,showkeys]{revtex4}
\usepackage[T1]{fontenc}
\usepackage[latin9]{inputenc}
\usepackage{amsmath}
\usepackage{amssymb}
\usepackage{graphicx}

\makeatletter
\@ifundefined{textcolor}{}
{%
 \definecolor{BLACK}{gray}{0}
 \definecolor{WHITE}{gray}{1}
 \definecolor{RED}{rgb}{1,0,0}
 \definecolor{GREEN}{rgb}{0,1,0}
 \definecolor{BLUE}{rgb}{0,0,1}
 \definecolor{CYAN}{cmyk}{1,0,0,0}
 \definecolor{MAGENTA}{cmyk}{0,1,0,0}
 \definecolor{YELLOW}{cmyk}{0,0,1,0}
 }

\usepackage{amsfonts}

\usepackage{times}

\usepackage{epstopdf}



\makeatother

\makeatother

\usepackage{babel}

\makeatother

\usepackage{babel}
\begin{document}

\title{Thermodynamics of coherent interfaces under mechanical stresses.
I. Theory}

\author{T. Frolov}

\email{tfrolov@gmu.edu}

\affiliation{Department of Physics and Astronomy, MSN 3F3, George Mason University,
Fairfax, Virginia 22030, USA}

\author{Y. Mishin}

\email{ymishin@gmu.edu}

\affiliation{Department of Physics and Astronomy, MSN 3F3, George Mason University,
Fairfax, Virginia 22030, USA}
\begin{abstract}
We present a thermodynamic theory of plane coherent solid-solid interfaces
in multicomponent systems subject to non-hydrostatic mechanical stresses.
The interstitial and substitutional chemical components are treated
separately using chemical potentials and diffusion potentials, respectively.
All interface excess quantities are derived using Cahn's (1979) generalized
excess method without resorting to geometric dividing surfaces. We
present expressions for the interface free energy as an excess quantity
and derive a generalized adsorption equation and an interface Gibbs-Helmholtz
equation which does not contain the interface entropy. The interface
stress tensor emerges naturally from the generalized adsorption equation
as an appropriate excess over bulk stresses and is shown to be generally
non-unique. Another interface property emerging from the generalized
adsorption equation is the interface excess shear. This property is
specific to coherent interfaces and represents the thermodynamic variable
conjugate to the shear stress applied parallel to the interface. The
theory reveals a number of Maxwell relations describing cross-effects
between thermal, chemical and mechanical responses of coherent interfaces.
In Part II of this work this theory will be applied to atomistic computer
simulations of grain boundaries.
\end{abstract}



\maketitle

\section{Introduction}

Thermodynamics properties of interfaces can have a strong impact on
microstructure development in materials by controlling phase nucleation,
growth, coarsening and many other processes.\cite{Balluffi95,Mishin2010a}.
The interface thermodynamics developed by Gibbs\cite{Willard_Gibbs}
was formulated in terms of interface excesses relative to an imaginary
geometric dividing surface separating the coexisting phases. Gibbs
defined the interface free energy $\gamma$ as the reversible work
expended to create a unit area of the interface. He showed that, while
other excess quantities generally depend on the choice of the dividing
surface, $\gamma$ is unique and thus a meaningful and measurable
quantity. Gibbs' work was focused on interfaces in fluid systems,
for which he derived the adsorption equation expressing the differential
$d\gamma$ in terms of differentials of temperature and chemical potentials
of the components present in the system. 

Gibbs\cite{Willard_Gibbs} also discussed solid-fluid interfaces and
pointed out that their interface area can change in two different
ways: when a new area of the interface is formed at fixed thermodynamic
states of the phases, and when the solid phase is elastically stretched
parallel to the interface. The second process leads to the definition
of the interface stress $\tau_{ij}$, a tensor quantity whose components
are generally different from $\gamma$ and can be positive or negative.\cite{Shuttleworth49}
Using a thought experiment with a solid equilibrated with three different
fluids, Gibbs demonstrated that chemical potential of a non-hydrostatically
stressed solid is not well-defined.\cite{Willard_Gibbs} At the time,
solid solutions were unknown and Gibbs considered only single-component
solids. When introducing $\gamma$ for solid-fluid interfaces, he
specifically placed the dividing surface so that the interface excess
of the solid component would vanish and there would be no need to
talk about its chemical potential. As was recently pointed out,\cite{Cammarata2009}
this approach would not work for a multicomponent solid.

Cahn\cite{Cahn79a} proposed a more general form of the adsorption
equation for hydrostatic systems by solving a system of Gibbs-Duhem
equations for the bulk phases and for a layer containing the interface.
By eliminating the Gibbsian construct of dividing surface, Cahn\textquoteright{}s
method offers a greater freedom of choice of intensive variables in
the adsorption equation. In particular, Cahn's formalism rigorously
introduces the interface excess volume, a quantity which is by definition
zero in Gibbs' thermodynamics. Cahn\cite{Cahn79a} also proposed a
Lagrangian ($L$) formulation of the Shuttleworth equation\cite{Shuttleworth49}
for phase boundaries, $\tau_{Lij}=\partial\gamma_{L}/\partial\varepsilon_{Lij}$,
and pointed to the importance of computing the derivative with respect
to the elastic strain $\varepsilon_{Lij}$ along a phase coexistence
path. For solid-fluid interfaces subject to non-hydrostatic mechanical
stresses, $\tau_{ij}$ has been formulated as an interface excess
quantity\cite{Frolov09a,Frolov09b,Frolov2010d,Frolov2010e} and computed
by atomistic methods for several crystallographic orientations.\cite{Frolov09a,Frolov09b,Frolov2010e,Frolov2010b} 

In comparison with solid-fluid interfaces, thermodynamics of solid-solid
interfaces is more challenging for at least two reasons. Firstly,
such interfaces are capable of supporting shear stresses parallel
to the interface plane. The interface response to such stresses depends
on the degree of coherency and can vary from perfect sliding for fully
incoherent interfaces to perfectly elastic response for fully coherent
interfaces.\cite{Robin1974,Larche73,Larche_Cahn_78,Larche1985} The
elastic response should obviously lead to additional terms in the
adsorption equation, with coefficients representing what can be called
``interface excess shears''. Such terms do not appear in existing
formulations of interface thermodynamics.\cite{Willard_Gibbs,Cahn79a}
Secondly, because of the undefined chemical potentials in non-hydrostatically
stressed solids, a different treatment is needed for the $-\Gamma_{i}d\mu_{i}$
terms appearing in the adsorption equation for fluid systems.\cite{Willard_Gibbs}
In the analysis of equilibrium between bulk solid phases, this problem
was circumvented by using chemical potentials for interstitial components
but diffusion potentials for substitutional components.\cite{Larche73,Larche_Cahn_78,Larche1985}
It will be shown below that the same approach can be transferred to
interface thermodynamics. 

Elastic response of coherent interfaces was also analyzed within mechanical
theories of interfaces.\cite{Gurtin-1975,Gurtin1998,Fried1999} In
such theories, the interface is treated as a surface separating two
elastic media subject to applied stresses. Mechanical equilibrium
conditions have been derived and possible excess deformations and
stresses at both plane and curved interfaces have been identified.
By contrast to thermodynamic theories,\cite{Robin1974,Larche73,Larche_Cahn_78,Larche1985}
the mechanical theories do not impose the chemical or phase equilibrium
conditions between the adjoining media. As a result, the interface
free energy $\gamma$ cannot be defined as the work of interface formation,
which blocks the route to the formulation of the adsorption equation.

In this paper we present a thermodynamic treatment of plane coherent
solid-solid interfaces subject to mechanical stresses. Our analysis
employs Cahn's generalized excess method\cite{Cahn79a} and the solid-solid
equilibrium theory developed by Robin\cite{Robin1974} and Larchè
and Cahn.\cite{Larche73,Larche_Cahn_78,Larche1985} As the authors
before us,\cite{Willard_Gibbs,Robin1974,Larche73,Larche_Cahn_78,Larche1985}
we do not rely on \emph{a priori} chosen thermodynamic potentials;
instead, all equations are derived \emph{directly} from the First
and Second Laws of thermodynamics. It is only after the derivation
is complete that some of the equations can be rewritten in simpler
and/or more intuitive forms by introducing appropriate thermodynamic
potentials. We start the paper by reviewing thermodynamics of a single
solid phase (Section \ref{sec:BTD}) and then formulate the coherent
phase equilibrium conditions (Section \ref{sec:Equilibrium-coherent})
in a form which prepares us for the subsequent thermodynamic analysis
of interfaces. Section \ref{sec:interface-thermodynamics} is central
to this paper. In it, we define the free energy $\gamma$ of a coherent
interface, reformulate it as an excess of appropriate thermodynamics
potentials, introduce a number of other interface excess quantities,
and finally derive the generalized adsorption equation and the interface
version of the Gibbs-Helmholtz equation. These equations identify
and define the interface excess shear, a property specific to coherent
interfaces and conjugate to the shear stress applied parallel to the
interface. They also define the interface stress tensor as an excess
quantity and demonstrate that it is not unique. We derive a number
of Maxwell relations describing interesting cross-effects between
different interface properties. In Section \ref{sec:other int} we
discuss how the proposed coherent interface theory can be applied
to incoherent interfaces and grain boundaries. Finally, in Section
\ref{sec:Discussion} we summarize our work and outline possible future
developments.

\section{Thermodynamics of a solid phase\label{sec:BTD}}

\subsection{The network solid}

Our treatment of a solid phase is based on the concept of a network
solid introduced by Robin\cite{Robin1974} and Larchè and Cahn.\cite{Larche73,Larche_Cahn_78}
We assume that the solid contains a penetrating network which is preserved
inside the solid and can be created or destroyed only at its boundaries.
The network serves three functions:
\begin{enumerate}
\item Enables a description of finite deformations of the solid\cite{Malvern69}
by associating physical points with network sites (or their small
groups).
\item Is capable of carrying mechanical loads, allowing the solid to reach
mechanical equilibrium under non-hydrostatic conditions.
\item Provides a \emph{conserved} set of sites, called substitutional, which
are completely, or almost completely, occupied by atoms.%
\footnote{For brevity we will be referring to the particles forming the solid
as atoms, although they can also be molecules as in the case of molecular
solids.%
} Accordingly, all chemical components can be divided into substitutional
(residing on substitutional sites) and interstitial (otherwise).%
\footnote{Some solids contain atoms capable of occupying both substitutional
and interstitial positions. This case is not discussed here but it
is straightforward to generalize our analysis to such solids.%
}
\end{enumerate}
In a crystalline solid, its lattice is formed by substitutional sites
and satisfies all three network properties. Since our theory is intended
primarily for applications to crystalline solids, we will adopt the
terminology in which we refer to the network as ``lattice'' and to
the network sites as ``lattice sites''. It should be noted, however,
that our results are of more general validity and do not require that
the solid have a long-range atomic ordering. The theory should be
equally applicable to non-periodic structures such as network glasses
or network polymers.

\subsection{Kinematics of deformation of a solid phase\label{sub:Kinematics}}

We will analyze the general case of \emph{finite} deformations of
a solid using the concept of a reference state.\cite{Malvern69} The
choice of the reference state is arbitrary, although it is often convenient
to choose a stress-free state. We will use the same Cartesian coordinate
system for both the reference and deformed states. For any physical
point defined by coordinates $x_{i}^{\prime}$ ($i=1,2,3$) in the
reference state, its coordinates $x_{i}$ in the deformed state are
functions of the reference coordinates, $\mathbf{x}=\mathbf{x}\left(\mathbf{x}^{\boldsymbol{\prime}}\right)$.
Any infinitesimal vector $dx_{i}$ connecting two physical points
in the deformed state is related to the infinitesimal vector $dx_{j}^{\prime}$
connecting the same two physical points in the reference state by
the linear transformation

\begin{equation}
dx_{i}={\displaystyle \sum_{j=1,2,3}}F_{ij}dx_{j}^{\prime},\label{eq:dx_Fdx}
\end{equation}
where tensor $\mathbf{F}$ is the deformation gradient with components

\begin{equation}
F_{ij}=\frac{\partial x_{i}}{\partial x_{j}^{\prime}}.\label{eq:F_def}
\end{equation}
It is assumed that $J:=\det\mathbf{F}\neq0$ and thus the reference
coordinates can be expressed as functions of the deformed ones, $\mathbf{x^{\boldsymbol{\prime}}}=\mathbf{x^{\boldsymbol{\prime}}}\left(\mathbf{x}\right)$.
The respective infinitesimal vectors are related by the inverse deformation
gradient $\mathbf{F^{\mathnormal{-1}}}$, 

\begin{equation}
dx_{i}^{\prime}={\displaystyle \sum_{j=1,2,3}}F_{ij}^{-1}dx_{j}.\label{eq:dxpr}
\end{equation}

\begin{figure}
\begin{centering}
\includegraphics[scale=0.7]{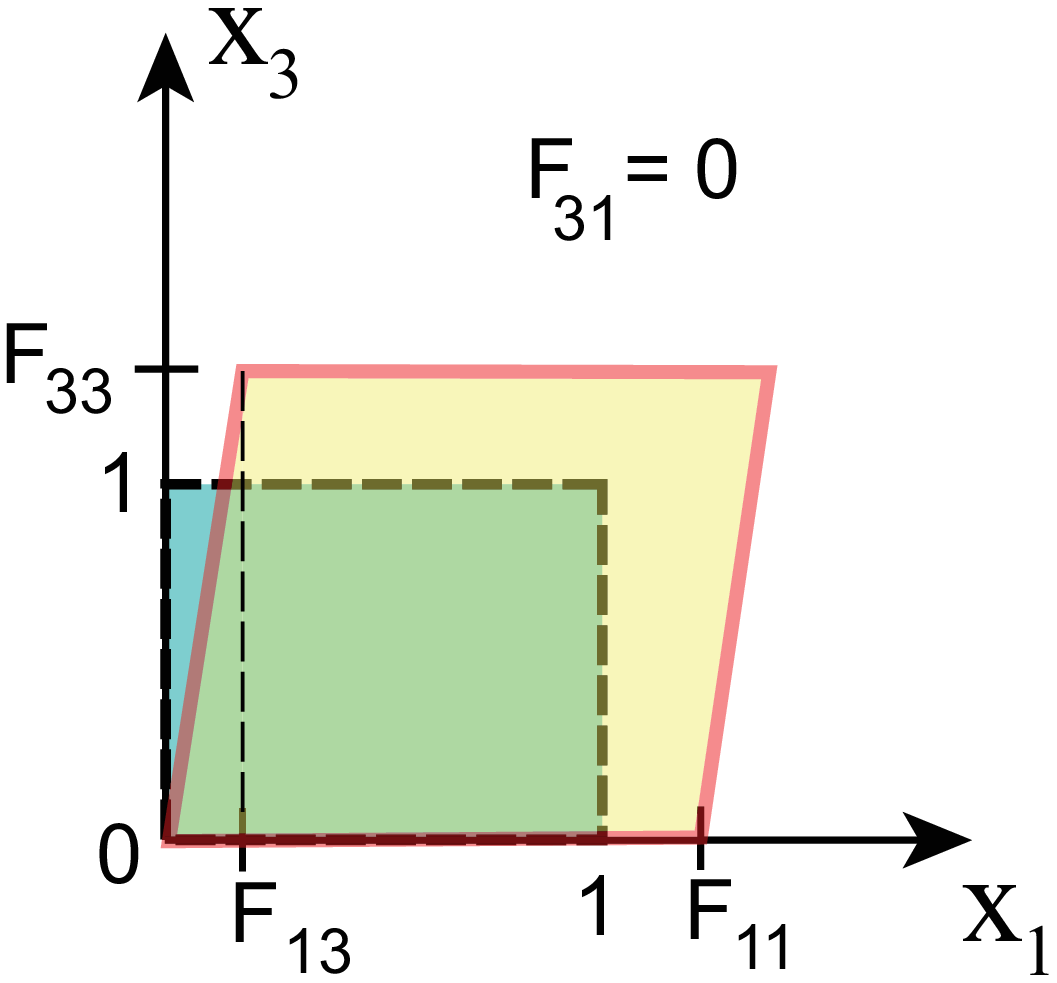}
\par\end{centering}

\caption{Two-dimensional schematic of a volume element undergoing a finite
deformation. In the reference state (dashed lines), the volume element
is a unit square. The components of the deformation gradient $\mathbf{F}$
correspond to the new lengths and projections of the edges of the
reference square in the deformed state (solid lines). \label{fig:F}}
\end{figure}

Only six components of $\mathbf{F}$ are needed to completely describe
all deformations (strains) of a solid. Without loss of generality,
we will set all sub-diagonal components of $\mathbf{F}$ to zero,

\begin{equation}
\mathbf{F}=\left(\begin{array}{ccc}
F_{11} & F_{12} & F_{13}\\
0 & F_{22} & F_{23}\\
0 & 0 & F_{33}
\end{array}\right),\label{eq:F_diag}
\end{equation}
with the Jacobian
\begin{equation}
J=F_{11}F_{22}F_{33}.\label{eq:A_detF}
\end{equation}
It can be shown that $\mathbf{F^{\mathnormal{-1}}}$ also has an upper-triangular
form with diagonal elements

\begin{equation}
F_{ii}^{-1}=1/F_{ii},\quad i=1,2,3.\label{eq:FF-1}
\end{equation}
The upper-triangular form of $\mathbf{F}$ implies that for any small
volume element with the shape of a parallelepiped, its bottom and
top faces remain normal to the $x_{3}$ axis. Furthermore, the edge
of the parallelepiped which is initially parallel to the $x_{1}$
axis remains parallel to it during the deformation. Deformation of
small volume element described by Eq.~(\ref{eq:F_diag}) is illustrated
by a two-dimensional schematic in Fig.~\ref{fig:F}.

\subsection{Thermodynamic description of a homogeneous solid phase\label{sub:Thermodynamics-solid}}

Consider a homogeneous multicomponent solid containing $K$ substitutional
and $L$ interstitial chemical components in a state of thermodynamic
equilibrium. The defining property of the substitutional components
is that their atoms fill a conserved set of lattice sites. Vacancies,
i.e. unoccupied substitutional sites, are neglected for the time being
and will be discussed separately in Section \ref{sec:Discussion}.
Thus, the total number $N$ of substitutional atoms in any given reference
region remains constant in all thermodynamic processes. Interstitial
atoms occupy otherwise empty positions between the lattice sites and
their number in any given reference region can vary. Diffusion of
both substitutional and interstitial atoms is allowed as long as it
preserves the substitutional sites. 

Consider a homogeneous region of the solid containing a total of $N$
substitutional and $n$ interstitial atoms and obtained by elastic
0deformation of a homogeneous reference region of volume $V^{\prime}$.
Suppose the reference region, and thus $V^{\prime}$ and $N$, are
fixed. Then the internal energy $U$ of the region is a function of
its entropy $S$, the amounts of individual chemical components $N_{k}$
and $n_{l}$, and the deformation gradient $\mathbf{F}$: 

\begin{equation}
U=U(S,N_{1},...,N_{K},n_{1},...,n_{L},\mathbf{F})\qquad\textnormal{(Fixed \ensuremath{N}).}\label{eq:U(S,N,F)}
\end{equation}
Due to the imposed substitutional constraint $\sum_{k}N_{k}=N=\textnormal{const}$,
only $(K-1)$ independent variations of $N_{k}$ are possible. To
implement this constraint, we can arbitrarily choose one of the substitutional
components as the reference component and assume that each time we
add to the solid an atom of a different substitutional component $k$,
we simultaneously remove an atom of the reference component.\cite{Larche73,Larche_Cahn_78,Larche1985}
Let us choose component $1$ as the reference and treat the amounts
of all other substitutional components as independent variables. The
amounts of the interstitial components can be varied without constraints.

Consider a reversible variation of state of the solid, with a fixed
reference region, when it exchanges heat with its environment, changes
its chemical composition, and performs mechanical work. The differential
of energy of the region is given by\cite{Larche73,Larche_Cahn_78}

\begin{equation}
dU=TdS+{\displaystyle \sum_{k=2}^{K}}M_{k1}dN_{k}+{\displaystyle \sum_{l=1}^{L}}\mu_{l}dn_{l}+\sum_{i,j=1,2,3}{\displaystyle V^{\prime}P_{ij}dF_{ji}},\label{eq:dU_V}
\end{equation}
where $T$ is temperature, $\mu_{l}$ are chemical potentials of the
interstitial atoms and $M_{k1}$ are $(K-1)$ diffusion potentials
of the substitutional atoms. According to Eq.~(\ref{eq:dU_V}), the
diffusion potential $M_{k1}$ is the energy change when an atom of
the substitutional component $k$ is replaced by an atom of the reference
component $1$ while keeping all other variables fixed:

\begin{equation}
M_{k1}=\frac{\partial U}{\partial N_{k}}-\frac{\partial U}{\partial N_{1}},\qquad k=2,\:...,\: K.\label{eq:M}
\end{equation}

In the last term in Eq.~(\ref{eq:dU_V}), $\mathbf{P}$ is the first
Piola-Kirchhoff stress tensor, which is generally not symmetrical
and is related to the symmetrical Cauchy stress tensor $\mathbf{\mathbf{\boldsymbol{\sigma}}}$
by\cite{Malvern69}

\begin{equation}
\mathbf{P}=J\mathbf{F^{\mathnormal{-1}}}\cdot\boldsymbol{\sigma}\label{eq:First PK tensor}
\end{equation}
(the dot denotes the inner product of tensors and vectors). Because
$\mathbf{\mathbf{F^{\mathnormal{-1}}}}$ is a upper-triangular matrix,
the components $P_{31}$, $P_{32}$ and $P_{33}$ are proportional
to the respective components of $\mathbf{\mathbf{\boldsymbol{\sigma}}}$: 

\begin{equation}
P_{3i}=F_{11}F_{22}\sigma_{3i}=\left(J/F_{33}\right)\sigma_{3i},\qquad i=1,2,3,\label{eq:PK_i3_C_i3}
\end{equation}
where we used Eqs.~(\ref{eq:F_diag}), (\ref{eq:A_detF}) and (\ref{eq:FF-1}). 

In preparation for the analysis of interfaces in Section \ref{sec:interface-thermodynamics},
we will rewrite the mechanical work term in Eq.~(\ref{eq:dU_V})
by separating the differentials $dF_{11}$, $dF_{12}$ and $dF_{22}$
from $dF_{13}$, $dF_{23}$ and $dF_{33}$:

\begin{equation}
dU=TdS+{\displaystyle \sum_{k=2}^{K}}M_{k1}dN_{k}+{\displaystyle \sum_{l=1}^{L}}\mu_{l}dn_{l}+\sum_{i=1,2,3}V^{\prime}F_{11}F_{22}\sigma_{3i}dF_{i3}+{\displaystyle \sum_{i,j=1,2}V^{\prime}}P_{ij}dF_{ji}.\label{eq:dU_prime_split}
\end{equation}
The $(K+L+6)$ differentials in the right-hand side of this equation
are independent and their number gives the total number of degrees
of freedom of a homogeneous solid phase.

\subsection{Relevant thermodynamic potentials}

Various thermodynamic potentials can be derived from Eq.~(\ref{eq:dU_prime_split})
by Legendre transformations. As will become clear later, the potential
relevant to coherent interfaces is
\begin{equation}
\Phi_{1}:=U-TS-{\displaystyle \sum_{k=2}^{K}}M_{k1}N_{k}-{\displaystyle \sum_{l=1}^{L}}\mu_{l}n_{l}-{\displaystyle \sum_{i=1,2,3}}\left(VF_{i3}/F_{33}\right)\sigma_{3i},\label{eq:Fi_N_def}
\end{equation}
where subscript 1 indicates the reference substitutional component
and $V=JV^{\prime}$ is the physical (deformed) volume of the homogeneous
solid. For a fixed reference volume (and thus $N$), 
\begin{equation}
\Phi_{1}=\Phi_{1}(T,M_{21},...,M_{K1},\mu_{1},...,\mu_{L},\sigma_{31},\sigma_{32},\sigma_{33},F_{11},F_{12},F_{22}).\label{eq:Fi1int}
\end{equation}
Using Eq.~(\ref{eq:dU_prime_split}) we obtain 
\begin{equation}
d\Phi_{1}=-SdT-{\displaystyle \sum_{k=2}^{K}N_{k}dM_{k1}}-{\displaystyle \sum_{l=1}^{L}n_{l}d\mu_{l}}-{\displaystyle \sum_{i=1,2,3}}\left(VF_{i3}/F_{33}\right)d\sigma_{3i}+{\displaystyle \sum_{i,j=1,2}}V^{\prime}Q_{ij}dF_{ji},\label{eq:dFi_N}
\end{equation}
where we denote

\begin{equation}
\mathbf{Q}:=J\mathbf{F}^{-1}\cdot\left(\mathbf{\boldsymbol{\sigma}}-{\displaystyle \sum_{m=1,2,3}}\frac{F_{m3}}{F_{33}}\sigma_{3m}\mathbf{I}\right)\label{eq:Q_hom}
\end{equation}
($\mathbf{I}\equiv\delta_{ij}$ is the identity tensor). Although
$\mathbf{Q}$ is a $3\times3$ tensor, only its components $Q_{11}$,
$Q_{21}$ and $Q_{22}$ appear in Eq.~(\ref{eq:dFi_N}). 

While the potential $\Phi_{1}$ will prove to be useful in interface
thermodynamics, its role in thermodynamics of a bulk phase is less
obvious. Unless the state of stress is hydrostatic, this potential
depends on the choice of the coordinate axes through the stress-strain
variables $\sigma_{31},\sigma_{32},\sigma_{33},F_{11},F_{12},F_{22}$.
In addition, $\Phi_{1}$ depends on the choice of the reference state
of strain. 

Eq.~(\ref{eq:Fi_N_def}) defines $\Phi_{1}$ for a homogeneous solid
region containing a given number $N$ of substitutional sites. We
can also define an intensive potential $\phi_{1}$ as $\Phi_{1}$
per substitutional atom:

\begin{equation}
\phi_{1}:=\Phi_{1}/N=U/N-TS/N-{\displaystyle \sum_{k=2}^{K}}M_{k1}C_{k}-{\displaystyle \sum_{l=1}^{L}}\mu_{l}c_{l}-{\displaystyle \sum_{i=1,2,3}}\left(\Omega F_{i3}/F_{33}\right)\sigma_{3i}.\label{eq:fi_def}
\end{equation}
Here $C_{k}:=N_{k}/N$ and $c_{l}:=n_{l}/N$ are concentrations of
substitutional and interstitial components per substitutional site,
$U/N$, $S/N$ and $\Omega$ are the energy, entropy and volume per
substitutional site, respectively. 

Similarly, we can introduce $K$ different potentials $\Phi_{m}$,
and accordingly $\phi_{m}$, by choosing other substitutional components
$m$ as the reference species:

\begin{equation}
\phi_{m}:=\Phi_{m}/N=U/N-TS/N-{\displaystyle \sum_{k=1}^{K}}M_{km}C_{k}-{\displaystyle \sum_{l=1}^{L}}\mu_{l}c_{l}-{\displaystyle \sum_{i=1,2,3}}\left(\Omega F_{i3}/F_{33}\right)\sigma_{3i}.\label{eq:fi2_def}
\end{equation}
Note that we have extended the summation with respect to $k$ from
1 to $K$ using the property $M_{kk}\equiv0$. Combining Eq.~(\ref{eq:fi2_def})
with known properties of diffusion potentials,\cite{Larche73,Larche_Cahn_78,Larche1985}
namely $M_{ik}=-M_{ki}$ and $M_{ij}=M_{ik}+M_{kj}$, the following
relationship between different $\phi$-potentials can be derived:

\begin{equation}
\phi_{m}-\phi_{n}=M_{mn},\qquad m,n=1,...K.\label{eq:f2_f1_M21}
\end{equation}
It also follows that

\begin{equation}
{\displaystyle \sum_{k=1}^{K}M_{km}C_{k}=\sum_{k=1}^{K}\left(\phi_{k}-\phi_{m}\right)C_{k}=\sum_{k=1}^{K}\phi_{k}C_{k}-\phi_{m}.}\label{eq:mc}
\end{equation}

Using Eqs.~(\ref{eq:fi2_def}) and (\ref{eq:mc}), we obtain the
following thermodynamic relation for a homogeneous non-hydrostatic
solid phase:

\begin{equation}
U-TS-{\displaystyle \sum_{i=1,2,3}}\left(VF_{i3}/F_{33}\right)\sigma_{3i}={\displaystyle \sum_{k=1}^{K}\phi_{k}N_{k}+\sum_{l=1}^{L}\mu_{l}}n_{l}.\label{eq:U_TS_PV_fi1_fi2}
\end{equation}
This equation closely resembles Gibbs' equation $U-TS+pV=\sum_{n}\mu_{n}N_{n}$
for hydrostatic systems ($p$ being external pressure),\cite{Willard_Gibbs}
with $\phi_{k}$ playing the role of chemical potentials. When the
solid is in a hydrostatic state of stress, $\sigma_{ij}=-\delta_{ij}p$,
the left-hand side of Eq.~(\ref{eq:U_TS_PV_fi1_fi2}) reduces to
the Gibbs free energy $U-TS+pV$. Accordingly, $\phi_{k}$ become
real chemical potentials of the substitutional components.

\subsection{The Gibbs-Duhem equation}

We can now derive a Gibbs-Duhem equation for a multicomponent non-hydrostatically
stressed solid. To this end, we again consider a variation of state
in which the solid region exchanges heat with its environment, performs
mechanical work, and changes its chemical composition by switching
chemical sorts of substitutional atoms (at fixed $N$) and changing
the amounts of interstitial atoms. Differentiating Eq.~(\ref{eq:Fi_N_def})
and using the relation $d\Phi_{1}=Nd\phi_{1}$ and $dU$ from Eq.~(\ref{eq:dU_prime_split}),
we obtain the following Gibbs-Duhem equation:

\begin{equation}
0=-SdT-{\displaystyle \sum_{k=2}^{K}}N_{k}dM_{k1}-Nd\phi_{1}-{\displaystyle \sum_{l=1}^{L}}n_{l}d\mu_{l}-{\displaystyle \sum_{i=1,2,3}}\left(VF_{i3}/F_{33}\right)d\sigma_{3i}+{\displaystyle \sum_{i,j=1,2}}V^{\prime}Q_{ij}dF_{ji}.\label{eq: Gibbs_Duhem}
\end{equation}
Applying Eq.~(\ref{eq:f2_f1_M21}), this equation can be rewritten
as 

\begin{equation}
0=-SdT-{\displaystyle \sum_{k=1}^{K}}N_{k}d\phi_{k}-{\displaystyle \sum_{l=1}^{L}}n_{l}d\mu_{l}-{\displaystyle \sum_{i=1,2,3}}\left(VF_{i3}/F_{33}\right)d\sigma_{3i}+{\displaystyle \sum_{i,j=1,2}}V^{\prime}Q_{ij}dF_{ji}.\label{eq: Gibbs_Duhem_in_fi}
\end{equation}

In the particular case of hydrostatic processes, $Q_{ij}\equiv0$
while $\sum_{i=1,2,3}\left(VF_{i3}/F_{33}\right)d\sigma_{3i}=-Vdp$.
In this case Eq.~(\ref{eq: Gibbs_Duhem_in_fi}) reduces to the classical
Gibbs-Duhem equation derived for fluids,\cite{Willard_Gibbs} 
\begin{equation}
0=-SdT-{\displaystyle \sum_{k=1}^{K+L}}N_{k}d\mu_{k}+Vdp,\label{eq:hydro-Gibbs-Duhem}
\end{equation}
where $\mu_{k}$ are chemical potentials of chemical components and
$N_{k}$ are their amounts.

Eq.~(\ref{eq: Gibbs_Duhem_in_fi}) is a relation between differentials
of the intensive variables which characterize thermodynamic states
of solids in equilibrium. By contrast to the standard Gibbs-Duhem
equation (\ref{eq:hydro-Gibbs-Duhem}), it contains non-hydrostatic
variations.

\section{Coherent equilibrium between solid phases\label{sec:Equilibrium-coherent}}

\subsection{Definition of coherency and coherent interface}

\begin{figure}
\centering{}\includegraphics[scale=0.6]{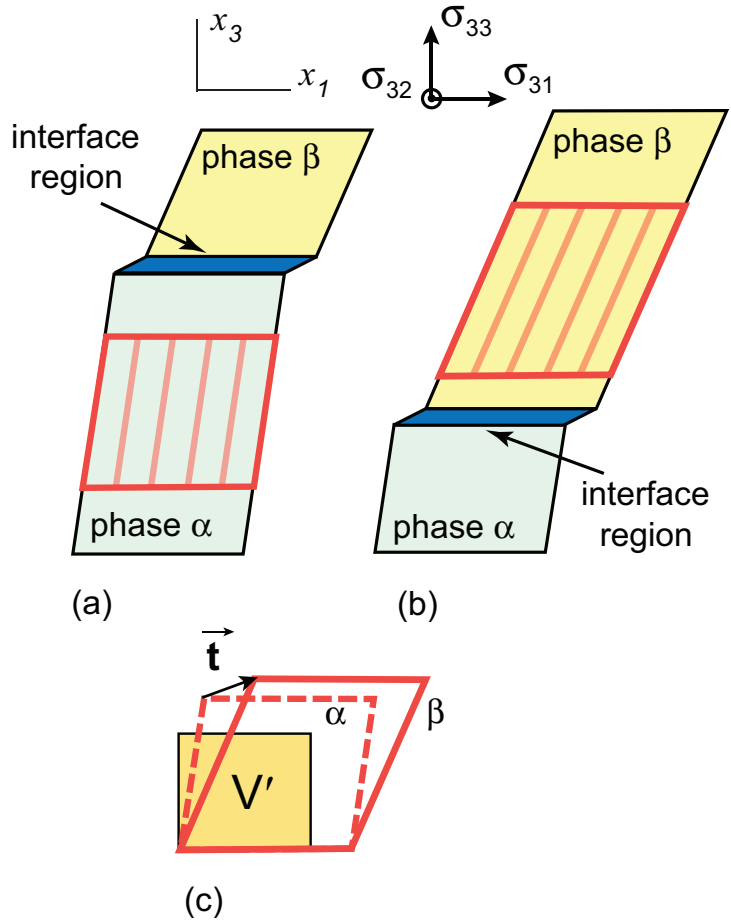}\caption{Two-dimensional schematic of phases $\alpha$ and $\beta$ separated
by a coherent interface. When the interface moves down, the striped
region of phase $\alpha$ shown in (a) transforms the striped region
of phase $\beta$ shown in (b). In (c), the reference volume element
$V^{\prime}$ (shaded unit square) transforms to the coherent phases
$\alpha$ (dashed lines) and $\beta$ (solid lines). The differences
between the deformation-gradient components $F_{13}$ and $F_{33}$
in the phases form the transformation vector $\mathbf{\boldsymbol{t}}$.
\label{fig:Phase-transformation-at}}
\end{figure}

We next discuss coherent equilibrium between two homogeneous solid
phases whose thermodynamic properties were introduced in Section \ref{sec:BTD}.
We assume that the two phases, which we refer to as $\alpha$ and
$\beta$, contain the same $K$ substitutional and $L$ interstitial
components and are separated by an infinitely large, plane coherent
interface normal to the $x_{3}$ direction (Fig.~\ref{fig:Phase-transformation-at}).
Our definition of phase coherency follows the works of Robin\cite{Robin1974}
and Larchè and Cahn.\cite{Larche73,Larche_Cahn_78,Larche1985} Namely,
a coherent transformation of a region of phase $\alpha$ to a region
of phase $\beta$ is accomplished by deformation of the lattice without
creation or destruction of lattice sites. Thus, a coherent transformation
fully preserves the reference region of the phase. All chemical components
are allowed to diffuse during the transformation as long as the lattice
sites remain intact. A more detailed discussion of the concept of
coherency and examples of coherent transformations can be found in
Refs.~\onlinecite{Larche73,Larche_Cahn_78,Larche1985,Robin1974}.

If a coherent transformation occurs on one side of a plane selected
inside a single-phase region, it produces a coherent interface between
the old and new phases. Advancement of the transformation front occurs
by interface migration. For coherent phases, there is a single network
of lattice sites penetrating through both phases and deformed during
the interface motion. In other words, the two-phase system can be
described as a deformation map of the same reference region as each
of the phases. Due to the lattice continuity across the interface,
sliding is prohibited and the two-phase system responds to applied
shear stresses elastically. (This is in contrast to incoherent interfaces,
which do not support static shear stresses and do not preserve the
lattice sites during their motion.) 

We will adopt the following kinematic description of coherent two-phase
systems. The deformation gradients of the phases, $\mathbf{F^{\alpha}}$
and $\mathbf{F^{\beta}}$, are taken relative to the \emph{same} reference
state and have the upper-triangular forms,

\begin{equation}
\mathbf{F^{\alpha}}=\left(\begin{array}{ccc}
F_{11} & F_{12} & F_{13}^{\alpha}\\
0 & F_{22} & F_{23}^{\alpha}\\
0 & 0 & F_{33}^{\alpha}
\end{array}\right),\label{eq:F_alpha}
\end{equation}

\begin{equation}
\mathbf{F^{\beta}}=\left(\begin{array}{ccc}
F_{11} & F_{12} & F_{13}^{\beta}\\
0 & F_{22} & F_{23}^{\beta}\\
0 & 0 & F_{33}^{\beta}
\end{array}\right),\label{eq:F_beta}
\end{equation}
where the superscripts indicate the phases. These forms ensure that
the $x_{3}$ direction in both phases remains normal to the interface
plane during all deformations. In addition, the lateral deformation
components $F_{11}$, $F_{12}$ and $F_{22}$ are common to both phases,
which is a necessary condition for the absence of sliding. Thus, the
two deformation gradients differ only in the components $F_{i3}.$
The differences between these components form a vector,
\begin{equation}
\mathbf{t}:=\left(F_{13}^{\beta}-F_{13}^{\alpha},F_{23}^{\beta}-F_{23}^{\alpha},F_{33}^{\beta}-F_{33}^{\alpha}\right),\label{eq:t-vector}
\end{equation}
which we call the \emph{transformation vector}. Its geometric meaning
is illustrated by the two-dimensional schematic in Fig.~\ref{fig:Phase-transformation-at}(c).

\subsection{Coherent phase equilibrium conditions\label{sub:Coherent-equilibrium}}

The conditions of coherent phase equilibrium were derived for a single-component
system by Robin\cite{Robin1974} and generalized to multicomponent
systems containing both substitutional and interstitial atoms by Larchè
and Cahn\cite{Larche73,Larche_Cahn_78} (see Voorhees and Johnson\cite{Voorhees2004}
for review). The equilibrium conditions can be summarized as follows:

(i) Temperature is uniform throughout the system. 

(ii) Diffusion potentials $M_{k1}$ of all substitutional components
and chemical potentials $\mu_{l}$ of all interstitial components
are uniform throughout the system. 

(iii) The internal mechanical equilibrium condition, $\mathbf{\nabla}^{\prime}\cdot\mathbf{P}=0$,
is satisfied inside each phase (the divergence is taken with respect
to the reference coordinates). 

(iv) The traction vector is continuous across the interface,
\begin{equation}
\mathbf{n^{\prime}}^{\alpha}\cdot\mathbf{P}^{\alpha}=-\mathbf{n^{\prime}}^{\beta}\cdot\mathbf{P}^{\beta},\label{eq:P_interf}
\end{equation}
where vectors $\mathbf{n}^{\prime\alpha}$ and $\mathbf{n^{\prime}}^{\beta}=-\mathbf{n}^{\prime\alpha}$
are unit normals to the phases in the reference state.%
\footnote{In their analysis of coherent equilibrium, Larchè and Cahn\cite{Larche73,Larche_Cahn_78}
used the first Piola-Kirchoff tensor which is the transpose of the
tensor used in our work. %
} This condition reflects the continuity of the displacement vector
across the interface. From Eqs.~(\ref{eq:P_interf}) and (\ref{eq:PK_i3_C_i3})
it follows that the Cauchy stress components $\sigma_{31}$, $\sigma_{32}$
and $\sigma_{33}$ are also continuous across the interface.

(v) Finally, the so-called phase-change equilibrium condition\cite{Larche73,Larche_Cahn_78}
must be satisfied. This condition expresses equilibrium with respect
to virtual displacements of the interface in which a layer of one
phase reversibly transforms to a layer of the other. Rewritten in
our notations, the phase change equilibrium condition derived by Larchè
and Cahn\cite{Larche73,Larche_Cahn_78} reads

\begin{equation}
\begin{array}{ccl}
 & U^{\alpha}-TS^{\alpha}-{\displaystyle \sum_{k=2}^{K}}M_{k1}N_{k}^{\alpha}-{\displaystyle \sum_{l=1}^{L}}\mu_{l}n_{l}^{\alpha}-{\displaystyle \sum_{i=1,2,3}}\left(VF_{i3}/F_{33}\right)^{\alpha}\sigma_{3i} & =\\
= & U^{\beta}-TS^{\beta}-{\displaystyle \sum_{k=2}^{K}}M_{k1}N_{k}^{\beta}-{\displaystyle \sum_{l=1}^{L}}\mu_{l}n_{l}^{\beta}-{\displaystyle \sum_{i=1,2,3}}\left(VF_{i3}/F_{33}\right)^{\beta}\sigma_{3i}.
\end{array}\label{eq:Phase_change}
\end{equation}
Here $U$, $S$, $V$ are the energy, entropy and volume of the phases
obtained by deformation of the same reference region. The total number
of substitutional atoms is equal in both phases, $N^{\alpha}=N^{\beta}$,
whereas the total number of interstitial components can be different
($n^{\alpha}\neq n^{\beta}$).

The equilibrium conditions (i)-(iii) are common to all types of interfaces.
The differences between the coherent, incoherent and other types of
interfaces lie in the remaining conditions (iv) and (v).

\subsection{Derivation of the phase-change equilibrium condition\label{sub:phase-change}}

The phase-change equilibrium condition (\ref{eq:Phase_change}) was
obtained from equation (41) of Larchè and Cahn\cite{Larche_Cahn_78}
by inserting our upper-triangular deformation gradients (\ref{eq:F_alpha})
and (\ref{eq:F_beta}) and the interface normal $\mathbf{n}=(0,0,1)$.
Note that Eq.~(\ref{eq:Phase_change}) contains the terms $\left(V^{\beta}F_{13}^{\beta}/F_{33}^{\beta}-V^{\alpha}F_{13}^{\alpha}/F_{33}^{\alpha}\right)\sigma_{31}$
and $\left(V^{\beta}F_{23}^{\beta}/F_{33}^{\beta}-V^{\alpha}F_{23}^{\alpha}/F_{33}^{\alpha}\right)\sigma_{32}$
proportional to the shear stresses $\sigma_{31}$ and $\sigma_{32}$.
These terms are specific to coherent interfaces and vanish for incoherent,
solid-fluid and fluid-fluid systems which do not support such stresses.
To elucidate the meaning of these terms and set the stage for the
analysis of interface thermodynamics, we will present an alternate
derivation of Eq.~(\ref{eq:Phase_change}) which assumes that the
equilibrium conditions (i) through (iv) are already satisfied. 

At fixed values of the intensive variables $T,M_{21},...,M_{K1},\mu_{1},...,\mu_{L},\sigma_{31},\sigma_{32},\sigma_{33},F_{11},F_{12},F_{22}$,%
\footnote{Out of this set of $(K+L+6)$ parameters, only $(K+L+5)$ are independent
(see Section \ref{sub:phase-coexist}). To maintain the neutral two-phase
equilibrium, only $(K+L+5)$ independent variables must be fixed,
which automatically fixes the remaining parameter.\label{fn:note-2}%
} equilibrium between the phases is neutral, i.e., the interface can
reversibly migrate up and down without altering thermodynamic states
of the bulk phases. The phase change equilibrium condition expresses
the neutrality of this equilibrium with respect to interface displacements.
Consider a homogeneous layer of phase $\alpha$ parallel to the interface
and containing a total of $N$ substitutional atoms. Suppose the interface
traveling down passes through this layer and transforms it completely
to a layer of phase $\beta$. The initial and transformed states of
the layer are shown schematically in Fig.~\ref{fig:Phase-transformation-at}(a,b).
In both states, the layer contains the same total number of substitutional
atoms, whereas the total number of interstitial atoms can be different.

Let us compute the change in internal energy of this layer. Because
the transformation is reversible, this change depends only on the
initial and final states (i.e., homogeneous phases $\alpha$ and $\beta$)
and not on the transformation path. As the interface traverses the
layer, it creates intermediate states that are not homogeneous. Instead
of examining this actual transformation process, we will consider
another, imaginary path on which the transformation occurs by homogeneous
deformation of the layer with a simultaneous change in its chemical
composition. Since the layer remains homogeneous during this process,
its energy change can be obtained by integrating Eq.~(\ref{eq:dU_prime_split})
derived previously for homogeneous variations. Remembering that the
intensive parameters are fixed, the integration gives

\begin{equation}
\begin{array}{ccl}
U^{\beta}-U^{\alpha}=T\left(S^{\beta}-S^{\alpha}\right)+{\displaystyle \sum_{k=2}^{K}}M_{k1}\left(N_{k}^{\beta}-N_{k}^{\alpha}\right)+{\displaystyle \sum_{l=1}^{L}}\mu_{l}\left(n_{l}^{\beta}-n_{l}^{\alpha}\right)+\\
+{\displaystyle \sum_{i=1,2,3}}\left(V^{\beta}F_{i3}^{\beta}/F_{33}^{\beta}-V^{\alpha}F_{i3}^{\alpha}/F_{33}^{\alpha}\right)\sigma_{3i}.
\end{array}\label{eq:deltaU_beta_alpha}
\end{equation}
The last term in Eq.~(\ref{eq:dU_prime_split}) does not contribute
to this equation because $F_{11}$, $F_{12}$ and $F_{22}$ are not
varied. Eq.~(\ref{eq:deltaU_beta_alpha}) recovers the phase-change
equilibrium condition (\ref{eq:Phase_change}).

This derivation emphasizes that the last term in Eq.~(\ref{eq:deltaU_beta_alpha})
represents the mechanical work, $W_{m}$, done by the stress components
$\sigma_{3i}$ during the phase transformation. This work term can
be rewritten as
\begin{equation}
{\displaystyle W_{m}=\sum_{i=1,2,3}}F_{11}F_{22}V^{\prime}\left(F_{i3}^{\beta}-F_{i3}^{\alpha}\right)\sigma_{3i}=F_{11}F_{22}V^{\prime}{\displaystyle \boldsymbol{\sigma}\cdot\mathbf{t}},\label{eq:work1}
\end{equation}
where $\mathbf{t}$ is the transformation vector defined by Eq.~(\ref{eq:t-vector})
and illustrated in Fig.~\ref{fig:Phase-transformation-at}(c). It
is important to note that, while $V^{\alpha}F_{i3}^{\alpha}/F_{33}^{\alpha}$
and $V^{\beta}F_{i3}^{\beta}/F_{33}^{\beta}$ individually depend
on the choice of the reference state of strain, vector $\mathbf{t}$
is an invariant and in principle measurable quantity characterizing
the geometry of the transformation.%
\footnote{See Section \ref{sec:Discussion} for a discussion of possible non-uniqueness
of the transformation strain and thus vector $\mathbf{t}$.%
} For incoherent and other interfaces incapable of supporting static
shear stresses, $W_{m}$ reduces to $F_{11}F_{22}V^{\prime}\sigma_{33}t_{3}=\left(V^{\beta}-V^{\alpha}\right)\sigma_{33}$.
For coherent interfaces, additional work is done by the shear stresses
along the components of $\mathbf{t}$ projected on the interface plane. 

Using the thermodynamic potential $\phi_{1}$ defined by Eq.~(\ref{eq:fi_def}),
the phase-change equilibrium condition (\ref{eq:Phase_change}) can
be formulated as simply $\phi_{1}^{\alpha}=\phi_{1}^{\beta}$. Furthermore,
by choosing other substitutional components as reference species,
the following $K$ relations can be obtained:

\begin{equation}
\phi_{m}^{\alpha}=\phi_{m}^{\beta}:=\phi_{m}\,,\qquad m=1,\,...,\, K.\label{eq:phase_change_fi}
\end{equation}
Thus, in a system with $K$ substitutional chemical components, there
are $K$ potentials that have the same value in coexisting phases.
This result resembles Gibbs' condition of equilibrium between fluid
phases,\cite{Willard_Gibbs} with $\phi_{m}$ playing the role of
chemical potentials.

\subsection{The equation of coherent phase coexistence in the parameter space\label{sub:phase-coexist}}

The Gibbs-Duhem equation (\ref{eq: Gibbs_Duhem}) establishes a relation
between the differentials of $(K+L+7)$ intensive parameters characterizing
a single-phase solid under stress. When two solid phases coexist,
their equilibrium imposes an additional constraint on possible variations
of state of the phases. This constraint can be formulated by writing
down the Gibbs-Duhem equation for each phase in terms of the same
set of intensive parameters and requiring that the two equations hold
simultaneously:

\begin{equation}
\begin{array}{ccl}
0 & = & -S^{\alpha}dT-{\displaystyle \sum_{k=2}^{K}}N_{k}^{\alpha}dM_{k1}-N^{\alpha}d\phi_{1}-{\displaystyle \sum_{l=1}^{L}}n_{l}^{\alpha}d\mu_{l}\\
 &  & -{\displaystyle \sum_{i=1,2,3}}\left(V^{\alpha}F_{i3}^{\alpha}/F_{33}^{\alpha}\right)d\sigma_{3i}+{\displaystyle \sum_{i,j=1,2}}V^{\prime\alpha}Q_{ij}^{\alpha}dF_{ji},
\end{array}\label{eq: GD_global_a}
\end{equation}

\begin{equation}
\begin{array}{ccl}
0 & = & -S^{\beta}dT-{\displaystyle \sum_{k=2}^{K}}N_{k}^{\beta}dM_{k1}-N^{\beta}d\phi_{1}-{\displaystyle \sum_{l=1}^{L}}n_{l}^{\beta}d\mu_{l}\\
 &  & -{\displaystyle \sum_{i=1,2,3}}\left(V^{\beta}F_{i3}^{\beta}/F_{33}^{\beta}\right)d\sigma_{3i}+{\displaystyle \sum_{i,j=1,2}}V^{\prime\beta}Q_{ij}^{\beta}dF_{ji}.
\end{array}\label{eq: GD_global_b}
\end{equation}
Note that these equations are written for arbitrarily chosen amounts
of the phases, i.e. generally $N^{\alpha}\neq N^{\beta}$. They can
be combined into one equation by eliminating one of the differentials.
This elimination leads to the equation

\begin{equation}
\begin{array}{ccl}
0 & = & -\{S\}_{X}dT-{\displaystyle \sum_{k=2}^{K}}\{N_{k}\}_{X}dM_{k1}-\{N\}_{X}d\phi_{1}-{\displaystyle \sum_{l=1}^{L}\{}n_{l}\}_{X}d\mu_{l}\\
 &  & -{\displaystyle \sum_{i=1,2,3}}\{VF_{i3}/F_{33}\}_{X}d\sigma_{3i}+{\displaystyle \sum_{i,j=1,2}}\{V^{\prime}Q_{ij}\}_{X}dF_{ji},
\end{array}\label{eq:bulk_coex}
\end{equation}
where $X$ is one of the extensive properties $S$, $N_{k}$ ($k=2,...,K$),
$N$, $n_{l}$ ($l=1,...,L$), $VF_{i3}/F_{33}$ ($i=1,2,3$) or $V^{\prime}Q_{ij}$
($i,j=1,2$). The curly braces are defined by 

\begin{equation}
\{Z\}_{X}:=Z^{\alpha}-Z^{\beta}X^{\alpha}/X^{\beta}\label{eq:bulk excess}
\end{equation}
for any pair of extensive properties $Z$ and $X$. The physical meaning
of $\{Z\}_{X}$ is the difference between the property $Z$ of the
two phases when they contain the same amount of $X$. For example,
$\{S\}_{N}$ is the difference between entropies of two homogeneous
regions of the phases containing the same total number of substitutional
atoms.

For any choice of $X$ out of the above list, the respective differential
coefficient in Eq.~(\ref{eq:bulk_coex}) vanishes because $\{X\}_{X}=0$.
The remaining $(K+L+6)$ terms form a differential equation defining
the coherent phase coexistence hypersurface in the configuration space
of intensive parameters. Thus, a system of two coexisting coherent
phases is capable of $(K+L+5)$ independent variations, which is one
degree of freedom less than for each phase taken separately. Knowing
one equilibrium state of the two-phase system, all other states can
be found by integrating Eq.~(\ref{eq:bulk_coex}) along different
paths on the phase coexistence hypersurface. 

Eq.~(\ref{eq:bulk_coex}) is an important result of this paper. It
provides the phase rule for equilibrium between coherent phases and
offers flexibility in choosing the independent variables corresponding
to the available degrees of freedom through the choice of $X$. It
generalizes the equation of phase coexistence derived by Gibbs for
solid-fluid interfaces\cite{Willard_Gibbs} by incorporating shears
parallel to the interface. Such shears are represented by the additional
terms $\{VF_{i3}/F_{33}\}_{X}d\sigma_{3i}$ with $i=1,2$. To further
elucidate the physical meaning of these terms, consider coherent equilibrium
between two binary substitutional solid solutions. For variations
of the shear stress at a constant temperature and fixed lateral dimensions
of the system,
\begin{equation}
\dfrac{dM_{21}}{d\sigma_{3i}}=-\dfrac{\{VF_{i3}/F_{33}\}_{N}}{\{N_{k}\}_{N}},\label{eq:bulk_coex_2}
\end{equation}
where we chose $X=N$. This relation predicts that to maintain the
equilibrium, variations in the diffusion potential in response to
variations in the shear stress must be proportional to the transformation
shear and inversely proportional to the difference between the phase
compositions. In other words, this relation describes changes in the
phase compositions caused by applied shear stresses.

It should be emphasized that Eq.~(\ref{eq:bulk_coex}) has been derived
under the assumption of interface coherency. One might think that
the phase coexistence equation for incoherent interfaces could be
obtained as simply a particular case of Eq.~(\ref{eq:bulk_coex})
when $\sigma_{31}$ and $\sigma_{32}$ are zero. This is not so. In
the absence of coherency, the lateral deformations of the phases ($F_{ij}^{\alpha}$
and $F_{ij}^{\beta}$, $i,j=1,2$) are not required to be equal and
can be varied independently. For example, one of the phases can be
stretched in a certain direction parallel to the incoherent interface
while the other compressed in the opposite direction. This deformation
produces interface sliding, which is a possible process for incoherent
interfaces. Furthermore, because the lattice sites can now be created
or destroyed when one phase transforms to the other, the deformation
gradients $\mathbf{F^{\alpha}}$ and $\mathbf{F^{\beta}}$ must be
defined relative to \emph{different} reference states. The incoherent
phase coexistence equation would have to be re-derived from the start,
which is beyond the scope of this paper.

\section{interface thermodynamics\label{sec:interface-thermodynamics}}

\subsection{The interface free energy $\gamma$\label{sub:gamma}}

We are now ready to analyze thermodynamics of coherent interfaces.
In this section we derive expressions for the interface free energy
$\gamma$ defined as the reversible work expended for creation of
a unit interface area. As above, we imagine two coexisting phases
$\alpha$ and $\beta$ separated by a coherent plane interface (Fig.~\ref{fig:Phase-transformation-at}),
but we now include the interface region as part of the system. Recall
that the deformation gradients $\mathbf{F^{\alpha}}$ and $\mathbf{F^{\beta}}$
were previously introduced for homogeneous phases and remain undefined
within the highly inhomogeneous interface region. We therefore need
to devise a method for introducing $\gamma$ and other interface excess
quantities without defining a deformation gradient inside the interface
region.

\begin{figure}
\noindent \begin{centering}
\includegraphics[clip,scale=0.7]{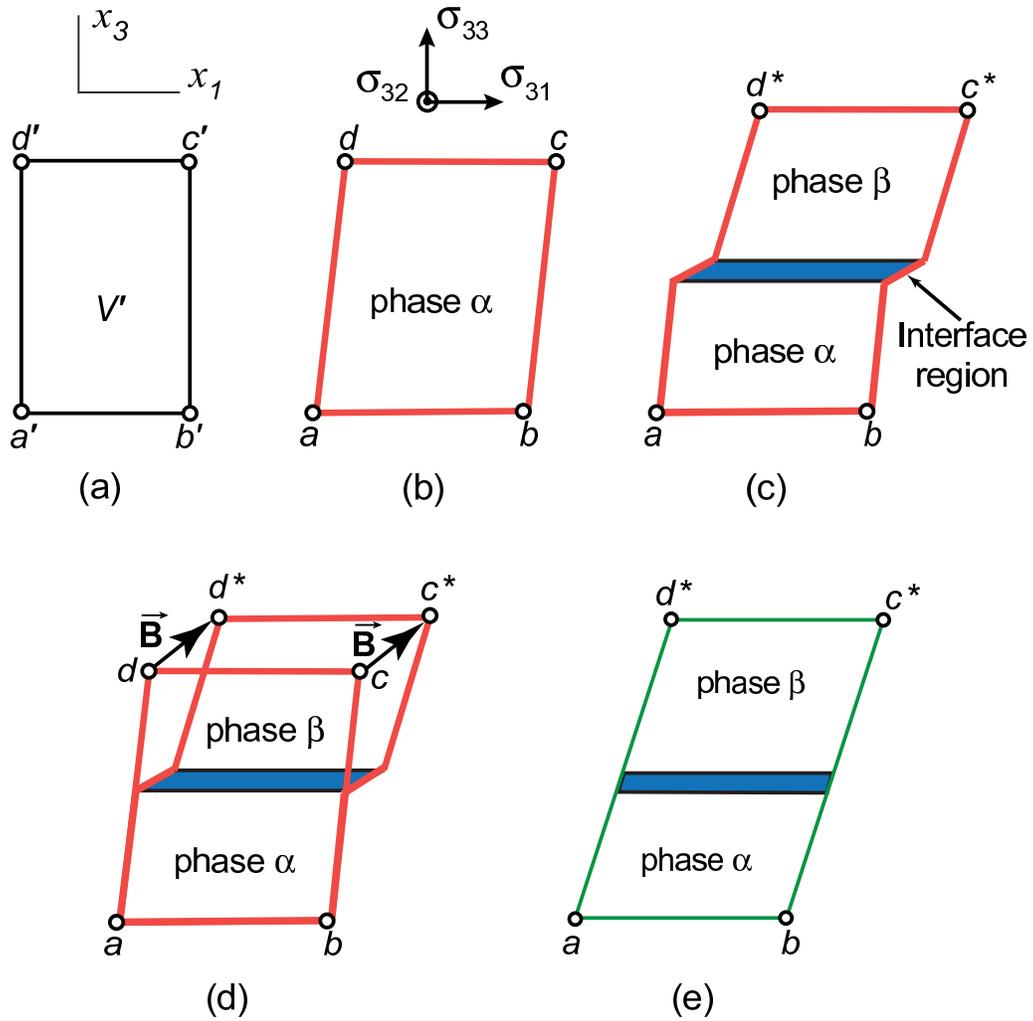}
\par\end{centering}

\caption{Two-dimensional schematic of coherent transformation of a region of
phase $\alpha$ to a two-phase region containing phases $\alpha$
and $\beta$ separated by an interface. (a) Reference state of the
region. (b) Deformed phase $\alpha$. (c) Two-phase region. (d) Overlapping
shapes of phase $\alpha$ and the two-phase region, showing the displacement
vector $\mathbf{B}$. The open circles mark selected physical points
labeled $a$ through $d$ with the prime indicating the reference
state and the asterisk indicating the two-phase state. The parallelogram
defined by the vertices $a$, $b$, $c^{*}$ and $d^{*}$ is shown
separately in (e).\label{fig:Creation-2}}
\end{figure}

As discussed earlier, the coherent two-phase equilibrium is neutral
when the intensive parameters $T,M_{21},...,M_{K1},\mu_{1},...,\mu_{L},\sigma_{31},\sigma_{32},\sigma_{33},F_{11},F_{12},F_{22}$
are fixed.%
\footnote{It will suffice to fix only $(K+L+5)$ of these $(K+L+6)$ parameters.
The remaining parameter is dependent and will be fixed automatically. %
} Consider a homogeneous region of phase $\alpha$ in the shape of
a parallelepiped with a reference volume $V^{\prime}$. Two faces
of the parallelepiped are parallel to the interface and one edge is
parallel to the $x_{1}$ axis. This is illustrated by a two-dimensional
schematic in Fig.~\ref{fig:Creation-2}(b), where the parallelepiped
is represented by a parallelogram. Suppose the interface spontaneously
migrates and enters this region, turning it into an equilibrium two-phase
system {[}Fig.~\ref{fig:Creation-2}(c){]}. Due to the coherency
condition, the cross-section of the region parallel to the interface
remains the same at every height. However, the shape of the region
changes due to the phase transformation strain. Consider a particular
position of the interface inside the two-phase region such that the
upper and lower boundaries of the region are deep inside the homogeneous
phases not perturbed by the presence of the interface. Suppose the
lower boundary of the region is fixed. Then the position of the upper
boundary generally changes as as result of the phase transformation.
Denote the displacement vector of the upper boundary $\mathbf{B}$. 

The geometric meaning of vector $\mathbf{B}$ is illustrated by the
two-dimensional schematic in Fig.~\ref{fig:Creation-2}. The initial
region $abcd$ is a deformation map of a reference region $a^{\prime}b{}^{\prime}c^{\prime}d{}^{\prime}$
with the deformation gradient $\mathbf{F^{\alpha}}$ {[}Fig.~\ref{fig:Creation-2}(a,b){]}.
After the upper part of the region transforms to phase $\beta$, it
becomes a map of the corresponding upper part of the reference region
with the deformation gradient $\mathbf{F^{\beta}}$. The reference
corners $c^{\prime}$ and $d{}^{\prime}$ are thus mapped to some
physical points $c^{*}$ and $d^{*}$ within the $\beta$ phase {[}Fig.~\ref{fig:Creation-2}(c){]}.
Vector $\mathbf{B}$ is defined as $cc^{*}$, or equivalently, $dd^{*}$
{[}Fig.~\ref{fig:Creation-2}(d){]}. Note that due to the conservation
of sites by coherent interfaces, the two-phase region contains the
same number $N$ of substitutional sites as the initial region of
phase $\alpha$. 

Vector $\mathbf{B}$ is used for calculation of the mechanical work
$W_{m}$ performed by stresses when the discussed region transforms
to the two-phase state. Since the cross-section of the region remains
fixed, the mechanical work is done only by the stress components $\sigma_{3i}$
when the upper boundary is displaced by vector $\mathbf{B}$. Thus,
$W_{m}=A\mathbf{n}^{\alpha}\mathbf{\boldsymbol{\cdot\sigma}}\cdot\mathbf{B}$,
where $\mathbf{n}^{\alpha}$ is the unit normal to the interface pointing
into phase $\beta$, $\mathbf{n}^{\alpha}\mathbf{\boldsymbol{\cdot\sigma}}$
is the traction vector, and $A$ is the cross-sectional area. 

To keep similarity with the mechanical work terms derived previously
for homogeneous phases (e.g., Eq.~(\ref{eq:deltaU_beta_alpha})),
we want to express $W_{m}$ through some deformation gradient. To
this end, we \emph{formally} define a \emph{homogeneous} deformation
gradient $\mathbf{\overline{F}}$ relative to the same reference state
as used for the homogeneous phases:
\begin{equation}
\mathbf{\overline{F}}:=\left(\begin{array}{ccc}
F_{11} & F_{12} & \left(F_{13}^{\alpha}+B_{1}A^{\prime}/V^{\prime}\right)\\
0 & F_{22} & \left(F_{23}^{\alpha}+B_{2}A^{\prime}/V^{\prime}\right)\\
0 & 0 & \left(F_{33}^{\alpha}+B_{3}A^{\prime}/V^{\prime}\right)
\end{array}\right),\label{eq:F-bar}
\end{equation}
where $A^{\prime}$ is the cross-sectional area of the interface in
the reference state. We will refer to $\mathbf{\overline{F}}$ as
the ``average'' deformation gradient of the region. The geometric
meaning of $\mathbf{\overline{F}}$ is the affine transformation that
carries the parallelepiped representing the reference region of phase
$\alpha$ to the parallelepiped formed by the corners of the two-phase
region after the phase transformation. In the two-dimensional schematic
shown in Fig.~\ref{fig:Creation-2}, $\mathbf{\overline{F}}$ transforms
the reference region $a^{\prime}b{}^{\prime}c^{\prime}d{}^{\prime}$
to the parallelogram $abc^{*}d^{*}$. The latter is shown separately
in Fig.~\ref{fig:Creation-2}(e). It should be noted that both $\mathbf{B}$
and $\mathbf{\overline{F}}$ generally depend on the choice of the
reference thickness $V^{\prime}/A^{\prime}$ of the $\alpha$ phase
region and on the position of the interface within the two-phase region.
In terms of $\mathbf{\overline{F}}$, the mechanical work term can
now be rewritten as 
\begin{equation}
W_{m}=A{\displaystyle \sum_{i=1,2,3}}\sigma_{3i}B_{i}={\displaystyle \sum_{i=1,2,3}}\left(V\overline{F}_{i3}/\overline{F}_{33}-V^{\alpha}F_{i3}^{\alpha}/F_{33}^{\alpha}\right)\sigma_{3i},\label{eq:Wm}
\end{equation}
where $V^{\alpha}=F_{11}F_{22}F_{33}^{\alpha}V^{\prime}$ and $V=F_{11}F_{22}\overline{F}_{33}V^{\prime}$
are physical volumes of the $\alpha$ phase region and the two-phase
region, respectively.%
\footnote{Because the two-phase region has the same cross-section at every height,
its volume equals the volume of the parallelepiped formed by its vertices,
see Fig.~\ref{fig:Creation-2}(e). The volume of this parallelepiped
is $F_{11}F_{22}\overline{F}_{33}V^{\prime}$.%
}

We next calculate the change in internal energy of the region when
it reversibly transforms from phase $\alpha$ to the two-phase state
{[}Fig.~\ref{fig:Creation-2}(b,c){]}. Instead of tracking the actual
motion of the interface into the region, we will consider only the
initial and final states and imagine a different reversible process
between them. Specifically, consider a process of homogeneous phase
transformation $\alpha\rightarrow\beta$ in the upper part of the
region at fixed $N$ and fixed intensive parameters $T,M_{21},...,M_{K1},\mu_{1},...,\mu_{L},\sigma_{31},\sigma_{32},\sigma_{33},F_{11},F_{12},F_{22}$.
Since the transformation occurs in an open system, its energy changes
due to the following processes: (i) heat exchange with the environment,
(ii) diffusion of atoms in and out of the system (at constant $N$),
(iii) mechanical work $W_{m}$ performed by stresses applied to the
boundaries of the region, and (iv) non-mechanical work $W_{nm}$ associated
with local atomic rearrangements leading to the formation of the interface.
Using Eq.~(\ref{eq:Wm}) for $W_{m}$, we have

\begin{equation}
\begin{array}{ccl}
U-U^{\alpha} & = & T\left(S-S^{\alpha}\right)+{\displaystyle \sum_{k=2}^{K}}M_{k1}\left(N_{k}-N_{k}^{\alpha}\right)+{\displaystyle \sum_{l=1}^{L}}\mu_{l}\left(n_{l}-n_{l}^{\alpha}\right)\\
 & + & {\displaystyle \sum_{i=1,2,3}}\left(V\overline{F}_{i3}/\overline{F}_{33}-V^{\alpha}F_{i3}^{\alpha}/F_{33}^{\alpha}\right)\sigma_{3i}+W_{nm},
\end{array}\label{eq:deltaU_gamma}
\end{equation}
where the extensive quantities with and without superscript $\alpha$
refer to the initial and final states, respectively. 

We define the interface free energy $\gamma$ as the non-mechanical
work done per unit interface area, i.e., $\gamma A:=W_{nm}$. Using
Eq.~(\ref{eq:fi_def}) for the $\alpha$ phase, Eq.~(\ref{eq:deltaU_gamma})
can be simplified to 
\begin{equation}
\gamma A=U-TS-{\displaystyle \sum_{k=2}^{K}}M_{k1}N_{k}-\phi_{1}N-{\displaystyle \sum_{l=1}^{L}}\mu_{l}n_{l}-{\displaystyle \sum_{i=1,2,3}}\left(V\overline{F}_{i3}/\overline{F}_{33}\right)\sigma_{3i},\label{eq:gamma_global}
\end{equation}
or expressing the diffusion potentials through the $\phi$-potentials
using Eq.~(\ref{eq:f2_f1_M21}),

\begin{equation}
\gamma A=U-TS-{\displaystyle \sum_{k=1}^{K}}\phi_{k}N_{k}-{\displaystyle \sum_{l=1}^{L}}\mu_{l}n_{l}-{\displaystyle \sum_{i=1,2,3}}\left(V\overline{F}_{i3}/\overline{F}_{33}\right)\sigma_{3i}.\label{eq:gamma_YM}
\end{equation}
These equations can be rewritten in a shorter form by introducing
the $\Phi_{1}$ potential of a two-phase region by analogy with Eq.~(\ref{eq:Fi_N_def}):

\begin{equation}
\Phi_{1}:=U-TS-{\displaystyle \sum_{k=2}^{K}}M_{k1}N_{k}-{\displaystyle \sum_{l=1}^{L}}\mu_{l}n_{l}-{\displaystyle \sum_{i=1,2,3}}\left(V\overline{F}_{i3}/\overline{F}_{33}\right)\sigma_{3i}.\label{eq:Fi-2}
\end{equation}
Then 

\begin{equation}
\gamma A=\Phi_{1}-N\phi_{1},\label{eq:gamma1}
\end{equation}
so that $\gamma$ is an excess of the $\Phi_{1}$ potential per unit
interface area. Of course, instead of component 1 we could have chosen
any other substitutional component as a reference. 

\begin{figure}
\noindent \begin{centering}
\includegraphics[scale=0.55]{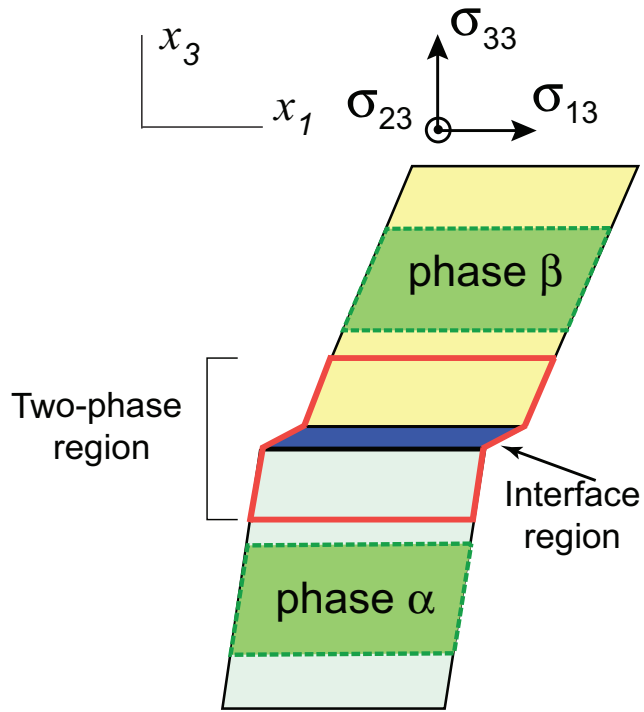}
\par\end{centering}

\caption{The two-phase region used in the derivation of expressions for the
total interface free energy $\gamma A$. The single-phase regions
used in the derivation are chosen outside the two-phase region. \label{fig:excess}}

\end{figure}

Eqs.~(\ref{eq:gamma_global}) and (\ref{eq:gamma_YM}) express the
total interface free energy $\gamma A$ through properties of an arbitrary
region containing the interface. While $\gamma A$ is uniquely defined
by these equations, the individual terms appearing in the right-hand
side depend on the location of the boundaries of the region. To express
these terms through interface excesses that are independent of the
boundaries, we need to subtract the contributions of the homogeneous
phases. To this end, we select two arbitrary regions inside the homogeneous
phases. Such single-phase regions can be chosen either inside or outside
the two-phase region. The latter case is illustrated in Fig.~\ref{fig:excess}.
Let the total numbers of substitutional atoms in the single-phase
regions be $N^{\alpha}$ and $N^{\beta}$, respectively (generally,
$N^{\alpha}$$\neq N^{\beta}$). Eq.~(\ref{eq:fi_def}) applied to
these regions gives

\begin{equation}
0=U^{\alpha}-TS^{\alpha}-{\displaystyle \sum_{k=2}^{K}}M_{k1}N_{k}^{\alpha}-\phi_{1}N^{\alpha}-{\displaystyle \sum_{l=1}^{L}}\mu_{l}n_{l}^{\alpha}-{\displaystyle \sum_{i=1,2,3}}\left(VF_{i3}/F_{33}\right)^{\alpha}\sigma_{3i}\label{eq:for_gamma_bulk_1}
\end{equation}
and

\begin{equation}
0=U^{\beta}-TS^{\beta}-{\displaystyle \sum_{k=2}^{K}}M_{k1}N_{k}^{\beta}-\phi_{1}N^{\beta}-{\displaystyle \sum_{l=1}^{L}}\mu_{l}n_{l}^{\beta}-{\displaystyle \sum_{i=1,2,3}}\left(VF_{i3}/F_{33}\right)^{\beta}\sigma_{3i}.\label{eq:for_gamma_bulk_b}
\end{equation}
Eqs.~(\ref{eq:gamma_global}), (\ref{eq:for_gamma_bulk_1}) and (\ref{eq:for_gamma_bulk_b})
form a system of three linear equations with respect to the same intensive
variables. We solve this system of equations for $\gamma A$ using
Cramer's rule of linear algebra.\cite{Frolov09b} The solution has
the form

\begin{equation}
\gamma A=[U]_{XY}-T[S]_{XY}-{\displaystyle \sum_{k=2}^{K}}M_{k1}[N_{k}]_{XY}-\phi_{1}[N]_{XY}-{\displaystyle \sum_{l=1}^{L}}\mu_{l}[n_{l}]_{XY}-{\displaystyle \sum_{i=1,2,3}}[V\overline{F}_{i3}/\overline{F}_{33}]_{XY}\sigma_{3i},\label{eq:gamma_excess}
\end{equation}
where $X$ and $Y\neq X$ are any two of the extensive quantities
$U$, $S$, $N_{k}$ ($k=2,...,K$), $N$, $n_{l}$ ($l=1,...,L$)
or $V\overline{F}_{i3}/\overline{F}_{33}$ ($i=1,2,3$). Note that
the last member of this list, corresponding to $i=3$, is simply volume
$V$. Using Eq.~(\ref{eq:f2_f1_M21}) we obtain the equivalent form
of $\gamma A$:

\begin{equation}
\gamma A=[U]_{XY}-T[S]_{XY}-{\displaystyle \sum_{k=1}^{K}}\phi_{k}[N_{k}]_{XY}-{\displaystyle \sum_{l=1}^{L}}\mu_{l}[n_{l}]_{XY}-{\displaystyle \sum_{i=1,2,3}}[V\overline{F}_{i3}/\overline{F}_{33}]_{XY}\sigma_{3i}.\label{eq:gamma_excess_fi}
\end{equation}
The coefficients $[Z]_{XY}$ are computed as ratios of two determinants:\cite{Cahn79a}

\begin{equation}
\left[Z\right]_{XY}:=\dfrac{\left\vert \begin{array}{ccc}
Z & X & Y\\
Z^{\alpha} & X^{\alpha} & Y^{\alpha}\\
Z^{\beta} & X^{\beta} & Y^{\beta}
\end{array}\right\vert }{\left\vert \begin{array}{cc}
X^{\alpha} & Y^{\alpha}\\
X^{\beta} & Y^{\beta}
\end{array}\right\vert }.\label{eq:def_determ}
\end{equation}
The quantities in the first row of the numerator are computed for
the region containing the interface, whereas all other quantities
are computed for arbitrary homogeneous regions of phases $\alpha$
and $\beta$. By properties of determinants,

\begin{equation}
[X]_{XY}=[Y]_{XY}=0,\label{eq:X_X_0}
\end{equation}
so that two terms in each of the Eqs.~(\ref{eq:gamma_excess}) and
(\ref{eq:gamma_excess_fi}) automatically vanish. 

The coefficient $[Z]_{XY}$ has the meaning of the interface excess
of extensive property $Z$ when the region containing the interface
contains the same amounts of $X$ and $Y$ as the two single-phase
regions combined; in other words, when the excesses of $X$ and $Y$
are zero. Thus, the excess of any property $Z$ is not unique; it
generally depends on the choice of the reference properties $X$ and
$Y$. If either $X$ or $Y$ is volume, then $[Z]_{XY}$ has the meaning
of the excess of $Z$ relative to a dividing surface similar to Gibbs'
formulation of interface thermodynamics.\cite{Willard_Gibbs}

The excesses $[N_{k}]_{XY}$ and $[n_{l}]_{XY}$ characterize the
segregated amounts of substitutional and interstitial components,
respectively. The terms $[V]_{XY}$, $[V\overline{F}_{13}/\overline{F}_{33}]_{XY}$
and $[V\overline{F}_{23}/\overline{F}_{33}]_{XY}$ define the excess
volume and two excess shears, respectively. For example, 
\begin{equation}
\left[V\overline{F}_{i3}/\overline{F}_{33}\right]_{SV}=\dfrac{\left\vert \begin{array}{ccc}
V\overline{F}_{i3}/\overline{F}_{33} & S & V\\
\left(VF_{i3}/F_{33}\right)^{\alpha} & S^{\alpha} & V^{\alpha}\\
\left(VF_{i3}/F_{33}\right)^{\beta} & S^{\beta} & V^{\beta}
\end{array}\right\vert }{\left\vert \begin{array}{cc}
S^{\alpha} & V^{\alpha}\\
S^{\beta} & V^{\beta}
\end{array}\right\vert },\qquad i=1,2.\label{eq:shears}
\end{equation}
In this case, the excess shears are taken with respect to a dividing
surface for which the excess of entropy is zero. The excess shears
are properties specific to coherent interfaces. They have no significance
for incoherent solid-solid interfaces, solid-fluid interfaces or any
other interfaces that cannot be equilibrated under applied shear stresses.
By contrast, the excess volume $[V]_{XY}$ is common to all types
of interfaces.\cite{Cahn79a} The numerical values of the excess volume
and excess shears depend on the choice of the reference properties
$X$ and $Y$. In the Gibbian formalism of dividing surface the excess
volume is zero by definition. 

The total interface free energy $\gamma A$ can be expressed through
excesses of different thermodynamic potentials corresponding to possible
choices of $X$ and $Y$. As already noted, $\gamma A$ can be expressed
as an excess of potential $\Phi_{1}$. Using our square bracket notation, 

\begin{equation}
\gamma A=[U-TS-{\displaystyle \sum_{k=2}^{K}}M_{k1}N_{k}-{\displaystyle \sum_{l=1}^{L}}\mu_{l}n_{l}-{\displaystyle \sum_{i=1,2}}\left(V\overline{F}_{i3}/\overline{F}_{33}\right)\sigma_{3i}]_{NV}\equiv[\Phi_{1}]_{NV},\label{eq:gamma_excess_N}
\end{equation}
i.e., the excess of $\Phi_{1}$ must be taken relative to the dividing
surface for which the excess of the total number of substitutional
atoms is zero. As another example,

\begin{equation}
\gamma A=[U-{\displaystyle \sum_{k=2}^{K}}M_{k1}N_{k}-{\displaystyle \sum_{l=1}^{L}}\mu_{l}n_{l}-{\displaystyle \sum_{i=1,2,3}}\left(V\overline{F}_{i3}/\overline{F}_{33}\right)\sigma_{3i}]_{NS},\label{eq:gamma_excess_N1}
\end{equation}
i.e., $\gamma A$ is an excesses of the potential appearing in the
square brackets when the excesses of the total number of substitutional
atoms and entropy are zero. The flexibility in expressing the same
quantity $\gamma A$ through excesses of different thermodynamic potentials
can be useful in applications of this formalism to experimental measurements
and simulations.

\subsection{The adsorption equation}

Having introduced the interface free energy, we are now in a position
to derive the generalized adsorption equation for coherent interfaces.
As the first step, we will compute the energy differential $dU$ for
a two-phase region containing the interface. We will take a region
in the shape of a parallelepiped as shown schematically in Fig.~\ref{fig:Creation-2}(e).
Recall that this shape is a map of the reference region of phase $\alpha$
containing the same number of substitutional atoms as in the parallelepiped.
This deformation map is formally defined by the deformation gradient
$\mathbf{\overline{F}}$ given by Eq.~(\ref{eq:F-bar}). Consider
a reversible variation in which this region exchanges heat and atoms
with its environment (at fixed $N$) and performs mechanical work
by elastically changing its shape and dimensions. The mechanical work,
$dW_{m}$, is done by the stresses applied to all faces of the parallelepiped
and equals the sum of the total forces exerted on the faces times
their displacements. The calculations give
\begin{equation}
dW_{m}=\sum_{i=1,2,3}V^{\prime}F_{11}F_{22}\sigma_{3i}d\overline{F}_{i3}+{\displaystyle \sum_{i,j=1,2}V^{\prime}}\overline{P}_{ij}dF_{ji},\label{eq:work-2}
\end{equation}
where $\overline{\mathbf{P}}:=\bar{J}\mathbf{\,\overline{F}^{\mathnormal{-1}}}\cdot\overline{\boldsymbol{\sigma}}$
is a formal analog of the first Piola-Kirchhoff stress tensor, $\overline{\boldsymbol{\sigma}}$
is the true stress tensor averaged over the volume of the parallelepiped,
$\bar{J}:=\det\mathbf{\mathbf{\overline{F}}}$, and $V^{\prime}$
is the reference volume of the phase $\alpha$ region. Because the
stress components $\sigma_{3i}$ are coordinate-independent and the
lateral stress components $\sigma_{ij}$ ($i,j=1,2$) depend only
on the coordinate $x_{3}$, it is only the lateral stress components
that must be averaged over $x_{3}$ in order to obtain $\overline{\boldsymbol{\sigma}}$.
Using the above expression for $dW_{m}$, the energy differential
equals

\begin{equation}
\begin{array}{ccl}
dU & = & TdS+{\displaystyle \sum_{k=2}^{K}}M_{k1}dN_{k}+{\displaystyle \sum_{l=1}^{L}}\mu_{l}dn_{l}\\
 & + & {\displaystyle \sum_{i=1,2,3}}V^{\prime}F_{11}F_{22}\sigma_{3i}d\overline{F}_{i3}+{\displaystyle \sum_{i,j=1,2}V^{\prime}}\overline{P}_{ij}dF_{ji}.
\end{array}\label{eq:dU_inhom}
\end{equation}
This equation looks similar to the previously derived Eq.~(\ref{eq:dU_prime_split})
and constitutes its generalization to inhomogeneous systems containing
a coherent interface.

At the next step, we take the differential of Eq.~(\ref{eq:gamma_global})
and insert $dU$ from Eq.~(\ref{eq:dU_inhom}). After some rearrangement
we obtain

\begin{eqnarray}
\begin{array}{ccl}
d\left(\gamma A\right) & = & -SdT-{\displaystyle \sum_{k=2}^{K}}N_{k}dM_{k1}-Nd\phi_{1}-{\displaystyle \sum_{l=1}^{L}}n_{l}d\mu_{l}\\
 & - & {\displaystyle \sum_{i=1,2,3}}\left(V\overline{F}_{i3}/\overline{F}_{33}\right)d\sigma_{3i}+{\displaystyle \sum_{i,j=1,2}}V^{\prime}\overline{Q}_{ij}dF_{ji}=d\Phi_{1}-Nd\phi_{1},
\end{array}\label{eq:dgammaA_global}
\end{eqnarray}
where we introduced the tensor 
\begin{equation}
\overline{\mathbf{Q}}:=\bar{J}\mathbf{\,\overline{F}^{\mathnormal{-1}}}\cdot\left(\overline{\boldsymbol{\sigma}}-{\displaystyle \sum_{m=1,2,3}}\frac{\overline{F}_{m3}}{\overline{F}_{33}}\sigma_{3m}\mathbf{I}\right).\label{eq:Q_layer}
\end{equation}
For a homogeneous phase, $\overline{\mathbf{Q}}$ reduces to the earlier
introduced tensor $\mathbf{Q}$, see Eq.~(\ref{eq:Q_hom}). 

The differentials in the right-hand side of Eq.~(\ref{eq:dgammaA_global})
are not independent. There are two constraints imposed by the Gibbs-Duhem
equations (\ref{eq: GD_global_a}) and (\ref{eq: GD_global_b}) containing
the same differentials. Solving the system of equations (\ref{eq:dgammaA_global}),
(\ref{eq: GD_global_a}) and (\ref{eq: GD_global_b}) by Cramer's
rule, we finally obtain the generalized adsorption equation

\begin{equation}
\begin{array}{ccl}
d\left(\gamma A\right) & = & -[S]_{XY}dT-{\displaystyle \sum_{k=2}^{K}}[N_{k}]_{XY}dM_{k1}-[N]_{XY}d\phi_{1}-{\displaystyle \sum_{l=1}^{L}[}n_{l}]_{XY}d\mu_{l}\\
 & - & {\displaystyle \sum_{i=1,2,3}}[V\overline{F}_{i3}/\overline{F}_{33}]_{XY}d\sigma_{3i}+{\displaystyle \sum_{i,j=1,2}}[V^{\prime}\overline{Q}_{ij}]_{XY}dF_{ji},
\end{array}\label{eq:the_AE}
\end{equation}
where $X$ and $Y$ are two of the extensive properties $S$, $N_{k}$
($k=2,...,K$), $N$, $n_{l}$ ($l=1,...,L$), $V\overline{F}_{i3}/\overline{F}_{33}$
($i=1,2,3$) or $V^{\prime}\overline{Q}_{ij}$ ($i,j=1,2$). 

Note the significant difference between Eqs.~(\ref{eq:dgammaA_global})
and (\ref{eq:the_AE}) written for the same differential $d(\gamma A)$.
In Eq.~(\ref{eq:dgammaA_global}), the differential coefficients
are properties of the entire region containing the interface. These
properties depend on the choice of the boundaries of the region and
thus have no physical significance. In the adsorption equation (\ref{eq:the_AE}),
on the other hand, the differential coefficients are interface excesses
$[Z]_{XY}$ defined by Eq.~(\ref{eq:def_determ}). For a given choice
of the reference properties $X$ and $Y$, such excesses are independent
of the boundaries of the region. Furthermore, the number of differentials
in the right-hand side of Eq.~(\ref{eq:dgammaA_global}) exceeds
the number $(K+L+5)$ of degrees of freedom of a coherent two-phase
system predicted by Eq.~(\ref{eq:bulk_coex}). By contrast, due to
the property (\ref{eq:X_X_0}) of determinants, two terms in Eq.~(\ref{eq:the_AE})
automatically vanish, leaving exactly $(K+L+5)$ independent differentials.
Each of the remaining excesses $[Z]_{XY}$ can be expressed as a partial
derivative of $\gamma A$ with respect to the corresponding intensive
variable and is therefore a measurable physical quantity. In terms
of the $\phi$-potentials, the adsorption equation takes the form

\begin{equation}
\begin{array}{ccl}
d\left(\gamma A\right) & = & -[S]_{XY}dT-{\displaystyle \sum_{k=1}^{K}}[N_{k}]_{XY}d\phi_{k}-{\displaystyle \sum_{l=1}^{L}[}n_{l}]_{XY}d\mu_{l}\\
 & - & {\displaystyle \sum_{i=1,2,3}}[V\overline{F}_{i3}/\overline{F}_{33}]_{XY}d\sigma_{3i}+{\displaystyle \sum_{i,j=1,2}}[V^{\prime}\overline{Q}_{ij}]_{XY}dF_{ji}.
\end{array}\label{eq:the_AE-1}
\end{equation}

The adsorption equation corresponding to Gibbs' formalism of the dividing
surface is obtained as a particular case of our adsorption equation
when either $X=V$ or $Y=V$. Although the excess volume $[V]_{XY}$
disappears, the excess shears $[V\overline{F}_{i3}/\overline{F}_{33}]_{XY}$
and $[V\overline{F}_{i3}/\overline{F}_{33}]_{XY}$ still remain. These
additional terms are not present in Gibbs' interface thermodynamics\cite{Willard_Gibbs}
or in Cahn's work.\cite{Cahn79a}

\subsection{The interface stress}

The terms in the adsorption equation that contain differentials of
the lateral deformation components $F_{11}$, $F_{12}$ and $F_{22}$
represent contributions to $\gamma A$ coming from elastic deformations
of the interface. These terms define the interface stress, the quantity
which was first discussed by Gibbs in the context of solid-fluid interfaces.\cite{Willard_Gibbs}

To formally define the interface stress tensor, choose the current
state of one of the phases as the reference state of strain. Then
$F_{11}=F_{22}=1$, $F_{12}=0$, and Eq.~(\ref{eq:the_AE}) becomes

\begin{equation}
\begin{array}{ccl}
d\left(\gamma A\right) & = & -[S]_{XY}dT-{\displaystyle \sum_{k=2}^{K}}[N_{k}]_{XY}dM_{k1}-[N]_{XY}d\phi_{1}-{\displaystyle \sum_{l=1}^{L}[}n_{l}]_{XY}d\mu_{l}\\
 & - & {\displaystyle \sum_{i=1,2,3}}[V\overline{F}_{i3}/\overline{F}_{33}]_{XY}d\sigma_{3i}+{\displaystyle \sum_{i,j=1,2}}\tau_{ij}^{XY}Ade_{ji},
\end{array}\label{eq:the_AE_e}
\end{equation}
where

\begin{equation}
\tau_{11}^{XY}:=\frac{1}{A}[V^{\prime}\overline{Q}_{11}]_{XY},\quad\tau_{22}^{XY}:=\frac{1}{A}[V^{\prime}\overline{Q}_{22}]_{XY},\quad\tau_{12}^{XY}=\tau_{21}^{XY}:=\frac{1}{A}[V^{\prime}\overline{Q}_{12}]_{XY}.\label{eq:tau_notat}
\end{equation}
In Eq.~(\ref{eq:the_AE_e}), $d\mathbf{e}$ is a $(2\times2)$ small-strain
tensor with components $de_{11}=dF_{11}$, $de_{22}=dF_{22}$ and
$de_{12}=de_{21}=\frac{1}{2}dF_{12}$. The $(2\times2)$ symmetrical
tensor $\boldsymbol{\mathbf{\mathbf{\mbox{\ensuremath{\tau}}}}}$
defined by Eqs.~(\ref{eq:tau_notat}) is the interface stress tensor
describing changes in the interface free energy due to its elastic
deformations. 

As other interface excess quantities, $\boldsymbol{\mathbf{\mathbf{\mbox{\ensuremath{\tau}}}}}$
generally depends on the choice of the reference properties $X$ and
$Y$ and is therefore not unique. However, in the particular case
when both phases are in a hydrostatic state of stress under a pressure
$p$, $\boldsymbol{\mathbf{\mathbf{\mbox{\ensuremath{\tau}}}}}$ becomes
independent of $X$ and $Y$ and is given by
\begin{equation}
\tau_{ij}=\dfrac{V}{A}\left(\overline{\sigma}_{ij}+p\delta_{ij}\right).\qquad i,j=1,2.\label{eq:tau-hydro}
\end{equation}
This equation immediately follows from the definition of the square
bracket $[V^{\prime}\overline{Q}_{ij}]_{XY}$ and the fact that for
hydrostatic phases $\overline{Q}_{ij}^{\alpha}=\overline{Q}_{ij}^{\beta}=0$.

Eqs.~(\ref{eq:tau_notat}) provide a recipe for interface stress
calculation when the phases are subject to non-hydrostatic stresses,
particularly when such stresses are different in the two phases (e.g.,
when one phase is under lateral tension while the other under lateral
compression). Previous calculations of interface stresses were focused
on unstressed or hydrostatically stressed phases. For solid-fluid
interfaces, the calculations for hydrostatic phases employed equations
similar to (\ref{eq:tau-hydro}). Non-hydrostatic stresses were included
only in surface stress calculations in single-phase systems.\cite{Frolov09a}
Calculations of $\boldsymbol{\mathbf{\mathbf{\mbox{\ensuremath{\tau}}}}}$
between non-hydrostatic solid phases using Eqs.~(\ref{eq:tau_notat})
is an uncharted territory and could be addressed in future work. 

In the remainder of this paper, the lateral deformations of a two-phase
system will be described by the small-strain tensor $d\mathbf{e}$
instead of the lateral components of the deformation gradient. As
mentioned above, this implies that the current state of one of the
phases is chosen as the reference state of strain. It should be emphasized
that (i) this assumption only reflects a particular choice of the
kinematic description of deformations, not a physical approximation,
and (ii) the normal and shear components $F_{i3}$ describing the
transformation strain between the two phases can still be finite.

\subsection{Lagrangian and physical forms of the adsorption equation}

Until this point we dealt with total excess quantities related to
the entire interface with an area $A$. It is often useful to define
\emph{specific} excesses, i.e. excesses per unit interface area in
either the current state or the reference state of strain. In the
former case, the excess quantity is referred to as physical while
in the latter case as Lagrangian. For example, $\gamma$ is the physical
specific excess of the interface free energy. The interface stress
defined by Eq.~(\ref{eq:tau_notat}) is the physical specific excess
of the tensor quantity $V^{\prime}\overline{Q}_{ij}$. One can also
define the Lagrangian interface free energy, $\gamma_{L}:=(\gamma A)/A^{\prime}$,
and the Lagrangian interface stress,
\begin{equation}
\tau_{L11}^{XY}:=\frac{1}{A^{\prime}}[V^{\prime}\overline{Q}_{11}]_{XY},\quad\tau_{L22}^{XY}:=\frac{1}{A^{\prime}}[V^{\prime}\overline{Q}_{22}]_{XY},\quad\tau_{L12}^{XY}=\tau_{L21}^{XY}:=\frac{1}{A^{\prime}}[V^{\prime}\overline{Q}_{12}]_{XY}.\label{eq:stress-L}
\end{equation}
Using Eq.~(\ref{eq:stress-L}) and the adsorption equation (\ref{eq:the_AE}),
we obtain the relation
\begin{equation}
\tau_{Lij}^{XY}=\left(\frac{\partial\gamma_{L}}{\partial e_{ij}}\right)^{XY}.\label{eq:Cahn}
\end{equation}
A similar relation involving the Lagrangian $\gamma$ was proposed
by Cahn.\cite{Cahn79a} Here, the superscript $XY$ in the right-hand
side indicates that the partial derivative is taken at fixed intensive
parameters, other than $e_{ij}$, that appear in the adsorption equation
when $X$ and $Y$ are chosen as the reference properties.

The Lagrangian form of the adsorption equation is obtained by dividing
Eq.~(\ref{eq:the_AE_e}) by $A^{\prime}$:
\begin{equation}
\begin{array}{ccl}
d\gamma_{L} & = & -\dfrac{[S]_{XY}}{A^{\prime}}dT-{\displaystyle \sum_{k=2}^{K}}\dfrac{[N_{k}]_{XY}}{A^{\prime}}dM_{k1}-\dfrac{[N]_{XY}}{A^{\prime}}d\phi_{1}-{\displaystyle \sum_{l=1}^{L}\dfrac{[n_{l}]_{XY}}{A^{\prime}}}d\mu_{l}\\
 & - & {\displaystyle \sum_{i=1,2,3}\dfrac{[V\overline{F}_{i3}/\overline{F}_{33}]_{XY}}{A^{\prime}}}d\sigma_{3i}+{\displaystyle \sum_{i,j=1,2}}\tau_{Lij}^{XY}de_{ji},
\end{array}\label{eq:the_AE_e-1}
\end{equation}
where the differential coefficients are Lagrangian specific excesses.
The physical form of the adsorption equation is obtained by differentiating
$\gamma A$ in Eq.~(\ref{eq:the_AE_e}) and using the relation $dA=A\sum_{i,j=1,2}\delta_{ij}de_{ij}$:

\begin{equation}
{\displaystyle \begin{array}{ccl}
d\gamma & = & -\dfrac{[S]_{XY}}{A}dT-{\displaystyle \sum_{k=2}^{K}}\dfrac{[N_{k}]_{XY}}{A}dM_{k1}-\dfrac{[N]_{XY}}{A}d\phi_{1}-{\displaystyle \sum_{l=1}^{L}\dfrac{[n_{l}]_{XY}}{A}}d\mu_{l}\\
 & - & {\displaystyle \sum_{i=1,2,3}\dfrac{[V\overline{F}_{i3}/\overline{F}_{33}]_{XY}}{A}}d\sigma_{3i}+{\displaystyle \sum_{i,j=1,2}}\left(\tau_{ij}^{XY}-\delta_{ij}\gamma\right)de_{ji}.
\end{array}}\label{eq:the_AE_phys}
\end{equation}
Now the differential coefficients give physical specific excesses.
From this equation, we immediately obtain the generalized form of
the Shuttleworth equation:\cite{Shuttleworth49}

\begin{equation}
\left(\frac{\partial\gamma}{\partial e_{ij}}\right)^{XY}=\tau_{ij}^{XY}-\delta_{ij}\gamma.\label{eq:Shuttle}
\end{equation}
The original Shuttleworth equation\cite{Shuttleworth49} was derived
for an open surface of a stress-free single-component solid deformed
isothermally. Eq.~(\ref{eq:Shuttle}) has been derived for coherent
interfaces in multicomponent systems in an arbitrary state of stress.
It actually represents a set of equations corresponding to different
choices of $X$ and $Y$ and thus different deformation paths.

Just as the Shuttleworth equation describes the effect of lateral
strains on the interface free energy, the following equations describe
the effect of the shear and normal stresses of $\gamma$ and $\gamma_{L}$
in the physical and Lagrangian forms, respectively:
\begin{equation}
\left(\frac{\partial\gamma}{\partial\sigma_{3i}}\right)^{XY}=\dfrac{[V\overline{F}_{i3}/\overline{F}_{33}]_{XY}}{A},\qquad i=1,2,3,\label{eq:shuttle-1}
\end{equation}
\begin{equation}
\left(\frac{\partial\gamma_{L}}{\partial\sigma_{3i}}\right)^{XY}=\dfrac{[V\overline{F}_{i3}/\overline{F}_{33}]_{XY}}{A^{\prime}},\qquad i=1,2,3.\label{eq:shuttle-2}
\end{equation}

\subsection{Thermodynamic integration}

We will now derive another version of the adsorption equation that
can be useful in applications. In principle, the interface free energy
$\gamma$ can be computed by integration of the adsorption equation
along a phase coexistence path knowing an initial value. However,
the excess entropy $[S]_{XY}$ appearing in this equation is rarely
accessible by experiments or simulations. To avoid calculation of
$[S]_{XY}$, we can eliminate it by combining Eqs.~(\ref{eq:gamma_excess})
and (\ref{eq:the_AE_e}) to obtain

\begin{eqnarray}
\begin{array}{ccl}
d{\displaystyle \left(\frac{\gamma A}{T}\right)} & = & {\displaystyle {\displaystyle -\frac{{\displaystyle \left[\Psi\right]}_{XY}}{T^{2}}}}dT-{\displaystyle {\displaystyle \sum_{k=2}^{K}}\frac{\left[N_{k}\right]_{XY}}{T}dM_{k1}}-{\displaystyle \frac{\left[N\right]_{XY}}{T}d\phi_{1}}-{\displaystyle \sum_{l=1}^{L}\frac{\left[n_{l}\right]_{XY}}{T}}d\mu_{l}\\
 & - & {\displaystyle {\displaystyle \sum_{i=1,2,3}}{\displaystyle \frac{\left[V\overline{F}_{i3}/\overline{F}_{33}\right]_{XY}}{T}}d\sigma_{3i}+{\displaystyle \frac{1}{T}}{\displaystyle \,\sum_{i,j=1,2}}\tau_{ij}^{XY}Ade_{ji}},
\end{array}\label{eq: GH}
\end{eqnarray}
where the thermodynamic potential $\Psi$ is defined by 
\begin{equation}
\Psi:=U-{\displaystyle \sum_{k=2}^{K}}N_{k}M_{k1}-N\phi_{1}-{\displaystyle \sum_{l=1}^{L}}n_{l}d\mu_{l}-{\displaystyle \sum_{i=1,2,3}}\sigma_{3i}V\overline{F}_{i3}/\overline{F}_{33}.\label{eq:PSI}
\end{equation}
It is straightforward to derive physical and Lagrangian forms of this
equation, whose left-hand sides will contain $d\left(\gamma/T\right)$
and $d\left(\gamma_{L}/T\right)$, respectively.

In the particular case when only temperature is varied, Eq.~(\ref{eq: GH})
gives
\begin{equation}
\left(\frac{\partial\left(\gamma A/T\right)}{\partial T}\right)^{XY}={\displaystyle -\frac{{\displaystyle \left[\Psi\right]}_{XY}}{T^{2}}}.\label{eq:GH-interface}
\end{equation}
This equation is similar to the classical Gibbs-Helmholtz equation\cite{Landau-Lifshitz-Stat-phys}
\begin{equation}
\left(\frac{\partial\left(G/T\right)}{\partial T}\right)_{p}={\displaystyle -\frac{U+pV}{T^{2}}}\label{eq:GH-bulk}
\end{equation}
for single-component fluid systems. Eq.~(\ref{eq: GH}) can be viewed
as a generalization of the Gibbs-Helmholtz equation to interfaces
in multicomponent systems. 

Eq.~(\ref{eq: GH}) can be used to compute $\gamma$ by integration
of $\gamma A/T$ along a trajectory on the phase coexistence hypersurface
in the configuration space of variables. The advantage of this integration
is that it does not require knowledge of $[S]_{XY}$. Free energy
calculations for solid-solid interfaces in multi-component elastically
stressed systems are presently non-existent and could be initiated
by applying the proposed thermodynamic integration approach.

\subsection{Maxwell relations\label{sub:Maxwell-relations}}

Because the adsorption equation contains the perfect differential
of $\gamma A$, it generates a number of Maxwell relations between
partial derivatives of the excess quantities. Similarly, the Gibbs-Helmholtz
equation (\ref{eq: GH}) is the perfect differential of $\gamma A/T$
and also generates Maxwell relations. We will focus on Maxwell relations
that involve the effects of mechanical stresses and strains on interface
properties. For hydrostatic precesses, such relations were discussed
by Cahn.\cite{Cahn79a} The additional terms in the adsorption equation
introduced in this work, such as the variations in the shear stresses
$\sigma_{31}$ and $\sigma_{32}$, lead to a number of additional
Maxwell relations. The Lagrangian and physical forms of the adsorption
equation produce different Maxwell relations, which will be presented
below side by side. In the partial derivatives appearing in these
relations, the variables which are held constant are dictated by the
particular choice of the extensive variables $X$ and $Y$. Thus,
each Maxwell relation actually represents a set of relations corresponding
to different choices of $X$ and $Y$.

\subsubsection{Mechanical relations}

The first set of Maxwell relations examines how the lateral deformations
$de_{ij}$ and the stresses $\sigma_{3k}$ affect the interface excess
volume, excess shears and interface stress. Using the Lagrangian and
physical forms of the adsorption equation, we obtain

\begin{equation}
\begin{array}{ccc}
{\displaystyle \frac{\partial\tau_{Lij}^{XY}}{\partial e_{kl}}=\frac{\partial\tau_{Lkl}^{XY}}{\partial e_{ij}},} &  & {\displaystyle \frac{\partial\left(\tau_{ij}^{XY}-\delta_{ij}\gamma\right)}{\partial e_{kl}}=\frac{\partial\left(\tau_{kl}^{XY}-\delta_{kl}\gamma\right)}{\partial e_{ij}}},\\
 &  & i,j,k,l=1,2,
\end{array}\label{eq:Max_E_E}
\end{equation}

\begin{equation}
\begin{array}{ccc}
{\displaystyle \frac{\partial\tau_{Lij}^{XY}}{\partial\sigma_{33}}=-\frac{\partial\left([V]_{XY}/A^{\prime}\right)}{\partial e_{ij}},} &  & {\displaystyle \frac{\partial\left(\tau_{ij}^{XY}-\delta_{ij}\gamma\right)}{\partial\sigma_{33}}=-\frac{\partial\left([V]_{XY}/A\right)}{\partial e_{ij}},}\\
 &  & i,j=1,2,
\end{array}\label{eq:Max_P_E}
\end{equation}

\begin{equation}
\begin{array}{ccc}
{\displaystyle \frac{\partial\tau_{Lij}^{XY}}{\partial\sigma_{3k}}=-\frac{\partial\left([VF_{k3}/F_{33}]_{XY}/A^{\prime}\right)}{\partial e_{ij}},} &  & {\displaystyle \frac{\partial\left(\tau_{ij}^{XY}-\delta_{ij}\gamma\right)}{\partial\sigma_{3k}}=-\frac{\partial\left([VF_{k3}/F_{33}]_{XY}/A\right)}{\partial e_{ij}},}\\
 &  & i,j,k=1,2,
\end{array}\label{eq:Max_s23_E}
\end{equation}

\begin{equation}
\frac{\partial\left([VF_{k3}/F_{33}]_{XY}/A^{\prime}\right)}{\partial\sigma_{33}}=\frac{\partial\left([V]_{XY}/A^{\prime}\right)}{\partial\sigma_{3k}},\quad k=1,2.\label{eq:Max_s33_s23}
\end{equation}
Eq.~(\ref{eq:Max_E_E}) represents the effect of lateral deformations
on the interface stress. Eq.~(\ref{eq:Max_P_E}) describes the interfacial
Poisson effect, in which lateral deformations of the interface produce
changes in the ``interface thickness'' (excess volume per unit area).
Because Eqs.~(\ref{eq:Max_E_E})-(\ref{eq:Max_s23_E}) involve changes
in interface area, their Lagrangian and physical forms are different.
By contrast, in Eq.~(\ref{eq:Max_s33_s23}) the derivatives are taken
at a constant area and thus the Lagrangian and physical forms are
identical. In such cases we present only the Lagrangian form of the
relation.

\subsubsection{Mechanochemical relations}

Elastic deformations parallel or normal to the interface affect interface
segregation. In turn, changes in segregation can produce changes in
interface stress, interface excess volume and interface excess shears.
We will present only Maxwell relations for substitutional components
when the diffusion potentials $M_{k1}$ are varied. For interstitial
components, the relations have a similar form but with the diffusion
potentials replaced by the chemical potentials $\mu_{l}$. The effect
of deformations parallel to the interface on the interface segregation
is described by the relations

\begin{equation}
\begin{array}{ccc}
{\displaystyle \frac{\partial\tau_{Lij}^{XY}}{\partial M_{k1}}=-\frac{\partial\left([N_{k}]_{XY}/A^{\prime}\right)}{\partial e_{ij}},} &  & {\displaystyle \frac{\partial\left(\tau_{ij}^{XY}-\delta_{ij}\gamma\right)}{\partial M_{k1}}=-\frac{\partial\left([N_{k}]_{XY}/A\right)}{\partial e_{ij}},}\\
 &  & i,j=1,2;\;\; k=2,...,K.
\end{array}\label{eq:Max_M12_E}
\end{equation}
Because the interface area changes, the Lagrangian and physical forms
of this relation are different. The effect of the stress components
$\sigma_{31}$, $\sigma_{32}$ and $\sigma_{33}$ on segregation is
described by the relations

\begin{equation}
\frac{\partial\left([V]_{XY}/A^{\prime}\right)}{\partial M_{k1}}=\frac{\partial\left([N_{k}]_{XY}/A^{\prime}\right)}{\partial\sigma_{33}},\qquad k=2,...,K,\label{eq:Max_M12_P}
\end{equation}

\begin{equation}
\begin{array}{ccc}
{\displaystyle \frac{\partial\left([VF_{i3}/F_{33}]_{XY}/A^{\prime}\right)}{\partial M_{k1}}=\frac{\partial\left([N_{k}]_{XY}/A^{\prime}\right)}{\partial\sigma_{3i}},} & \qquad i=1,2;\;\; k=2,...,K. & {\displaystyle }\end{array}\label{eq:Max_M12_s23}
\end{equation}
Since the derivatives are taken at a constant area, the Lagrangian
and physical forms are identical.

\subsubsection{Thermomechanical relations}

Such relations describe the effects of temperature on interface stress,
excess volume and excess shears. The relations generated by the adsorption
equation would contain the excess entropy $[S]_{XY}$ which is not
easily accessible. Instead, we will use the Gibbs-Helmholtz equation
(\ref{eq: GH}) which does not contain $[S]_{XY}$. The following
Maxwell relations are obtained: 

\begin{equation}
\begin{array}{ccc}
{\displaystyle \frac{\partial\left(\tau_{Lij}^{XY}/T\right)}{\partial T}=-\frac{\partial\left([\Psi]_{XY}/A^{\prime}T^{2}\right)}{\partial e_{ij}},} &  & {\displaystyle \frac{\partial\left\{ \left(\tau_{ij}^{XY}-\delta_{ij}\gamma\right)/T\right\} }{\partial T}=-\frac{\partial\left([\Psi]_{XY}/AT^{2}\right)}{\partial e_{ij}},}\\
 &  & i,j=1,2,
\end{array}\label{eq:Max_U_T_E}
\end{equation}

\begin{equation}
\frac{\partial\left([V]_{XY}/A^{\prime}T\right)}{\partial T}=\frac{\partial\left([\Psi]_{XY}/A^{\prime}T^{2}\right)}{\partial\sigma_{33}},\label{eq:Max_U_T_P}
\end{equation}

\begin{equation}
\frac{\partial\left([VF_{k3}/F_{33}]_{XY}/A^{\prime}T\right)}{\partial T}=\frac{\partial\left([\Psi]_{XY}/A^{\prime}T^{2}\right)}{\partial\sigma_{3k}},\qquad k=1,2.\label{eq:Max_U_T_s23}
\end{equation}
where the potential $\Psi$ is given by Eq.~(\ref{eq:PSI}). In Eq.~(\ref{eq:Max_U_T_E}),
the Lagrangian and physical forms of the same relation are different.

\subsubsection{Thermochemical relations}

Using the Gibbs-Helmholtz equation (\ref{eq: GH}), we can evaluate
the effect of temperature on interface segregation of substitutional
and interstitial components. The corresponding derivatives involve
the excess $[\Psi]_{XY}$ instead of $[S]_{XY}$ and read

\begin{equation}
\frac{\partial\left([N_{k}]_{XY}/A^{\prime}T\right)}{\partial T}=\frac{\partial\left([\Psi]_{XY}/A^{\prime}T^{2}\right)}{\partial M_{k1}},\qquad k=2,...,K,\label{eq:Max_U_T_M12}
\end{equation}

\begin{equation}
\frac{\partial\left([n_{l}]_{XY}/A^{\prime}T\right)}{\partial T}=\frac{\partial\left([\Psi]_{XY}/A^{\prime}T^{2}\right)}{\partial\mu_{l}}.\qquad l=1,...,L.\label{eq:Max_U_mu_l}
\end{equation}
For substitutional components, the derivatives are taken with respect
to diffusion potentials, whereas for interstitial components with
respect to chemical potentials. The Lagrangian and physical forms
of these relations are identical.

\section{Relation to other types of interfaces\label{sec:other int} }

\subsection{Incoherent solid-solid interfaces}

Incoherent solid-solid interfaces differ from coherent in two ways: 
\begin{enumerate}
\item Lateral deformations of the two phases are allowed to be different
and independent of each other, as long as they preserve the orientation
of the interface plane. Accordingly, the deformation gradients of
the phases must still have the upper-triangular form but need not
satisfy Eqs.~(\ref{eq:F_alpha}) and (\ref{eq:F_beta}) with equal
lateral components.
\item Incoherent interfaces do not support static shear stresses applied
parallel to the interface plane, responding to such stresses by sliding. 
\end{enumerate}
These conditions can be reconciled with the coherent interface theory
by considering only processes in which the shear stresses $\sigma_{31}$
and $\sigma_{32}$ are identically zero and the lateral deformations
of the phases remain equal. Under these constraints, the incoherency
of the interface does not manifest itself and all equations derived
for coherent interfaces are valid for incoherent ones. The number
of independent variables reduces to $(K+L+3)$ and all equations are
significantly simplified. In particular, the equations no longer contain
the components $\overline{F}_{i3}$ of the deformation gradient $\mathbf{\overline{F}}$
{[}Eq.~(\ref{eq:F-bar}){]} and the latter need not be introduced.
The shapes of the two-phase and single-phase regions used in the thought
experiments become unimportant; only their volumes appear in the final
equations. 

For example, the total interface free energy $\gamma A$ is given
by the simplified form of Eq.~(\ref{eq:gamma_excess_fi}), 
\begin{equation}
\gamma A=[U]_{XY}-T[S]_{XY}-{\displaystyle \sum_{k=1}^{K}}\phi_{k}[N_{k}]_{XY}-{\displaystyle \sum_{l=1}^{L}}\mu_{l}[n_{l}]_{XY}-[V]_{XY}\sigma_{33}.\label{eq:gamma_excess_fi-1}
\end{equation}
The potentials $\phi_{m}$ defined by Eq.~(\ref{eq:fi2_def}) reduce
to

\begin{equation}
\phi_{m}=U/N-TS/N-{\displaystyle \sum_{k=1}^{K}}M_{km}C_{k}-{\displaystyle \sum_{l=1}^{L}}\mu_{l}c_{l}-\sigma_{33}\Omega.\label{eq:fi2_def-1}
\end{equation}
The adsorption equation becomes
\begin{equation}
d\left(\gamma A\right)=-[S]_{XY}dT-{\displaystyle \sum_{k=1}^{K}}[N_{k}]_{XY}d\phi_{k}-{\displaystyle \sum_{l=1}^{L}[}n_{l}]_{XY}d\mu_{l}-[V]_{XY}d\sigma_{33}+{\displaystyle \sum_{i,j=1,2}}\tau_{ij}^{XY}Ade_{ji},\label{eq:AE-2}
\end{equation}
where we use the current state of one of the phases as the reference
state of lateral strain. The interface stress tensor simplifies to
\begin{equation}
\tau_{ij}^{XY}=\frac{1}{A}\left[V\left(\overline{\sigma}_{ij}-\sigma_{33}\delta_{ij}\right)\right]_{XY},\label{eq:incoh-interf-stress}
\end{equation}
where $\overline{\sigma}_{ij}$ is the Cauchy stress tensor averaged
over a region of volume $V$.

It should be emphasized, however, that the above equations describe
only some of the possible state variations of an incoherent two-phase
systems. They do not include variations in which the phases undergo
different lateral deformations and thus slip against each other. Due
to such variations, an incoherent two-phase system possesses more
degrees of freedom than a coherent one with the same number of substitutional
and interstitial components. Thus, incoherent interfaces cannot be
considered a particular case of coherent interfaces. They require
a separate treatment, which will be presented elsewhere.

\subsection{Grain boundaries}

Grain boundary is an interface between regions of the same crystalline
phase with different lattice orientations. As other solid-solid interfaces,
GBs can be coherent or incoherent. Coherent GBs can support not only
stresses normal to the GB plane but also shear stresses parallel to
it. When temperature\cite{Broughton86a} and/or chemical composition\cite{Williams09}
change, some coherent GBs can change their structure to one that permits
GB sliding. The GB becomes incoherent. 

Two different cases must be distinguished: when the grains are thermodynamically
identical and when they are not. By definition, the grains are considered
thermodynamically identical when the phase-change equilibrium condition
(\ref{eq:deltaU_beta_alpha}) is satisfied as a mathematical identity
once the equilibrium conditions (i)-(iv) formulated in Section \ref{sub:Coherent-equilibrium}
are satisfied. In other words, the phase-change equilibrium condition
need not be imposed as a separate equation of constraint. This can
be the case when the grains are stress-free and uninfluenced by electric,
magnetic or other fields. Thermodynamically identical grains can be
treated as parts of the same single-phase system. On the other hand,
in the presence of mechanical stresses or applied fields, the equilibrium
thermodynamic states of the grains can be different. For example,
when the solid is elastically anisotropic and the grains are subject
to mechanical stresses, they either never reach equilibrium or can
reach an equilibrium state in which their elastic strain energy densities
and chemical compositions are different. In the latter case, the phase-change
equilibrium condition (\ref{eq:deltaU_beta_alpha}) is not satisfied
automatically and must be imposed as a separate constraint. Such cases
should be formally treated as if the grains were two different phases.
Accordingly, all thermodynamic equations developed in Sections \ref{sec:Equilibrium-coherent}
and \ref{sec:interface-thermodynamics} for phase boundaries directly
apply to this case.

There are situations when, due to crystal symmetry, the grains remain
thermodynamically identical even in the presence of certain mechanical
stresses. As an example, consider a coherent symmetrical tilt GB.
In the unstressed state, the grains are identical and form a single-phase
system. Due to the mirror symmetry across the boundary plane, the
lateral deformations $de_{ij}$ ($i,j=1,2$) and the normal stress
$\sigma_{33}$ leave the grains identical. Moreover, due to the twofold
symmetry around the axis $x_{2}$ normal to the tilt axis, the shear
stress $\sigma_{31}$ also leaves the grains identical. Thus, when
the system is subject to these deformations, it continues to be a
single-phase system.%
\footnote{By contrast, the shear stress $\sigma_{32}$ applied normal to the
tilt axis destroys the identity of the grains. This stress causes
coupled GB motion,\cite{Cahn06b} which cannot be prevented unless
we create different chemical compositions in the grains and thus a
thermodynamic driving force balancing the driving force of coupled
motion. But then the grains essentially become two different phases.%
} That Eq.~(\ref{eq:deltaU_beta_alpha}) is satisfied in this case
as an identity can be seen from the fact that the differences $\left(U^{\beta}-U^{\alpha}\right)$,
$\left(S^{\beta}-S^{\alpha}\right)$, $\left(N_{k}^{\beta}-N_{k}^{\alpha}\right)$
and $\left(n_{l}^{\beta}-n_{l}^{\alpha}\right)$ related to grain
regions containing the same total numbers of substitutional atoms
are zero by the symmetry. The remaining terms in Eq.~(\ref{eq:deltaU_beta_alpha})
represent the mechanical work $W_{m}$ and are given by Eq.~(\ref{eq:work1}).
For a symmetrical tilt boundary $F_{13}^{\beta}=F_{13}^{\alpha}$
and $F_{33}^{\beta}=F_{33}^{\alpha}$; only the shear components $F_{23}^{\alpha}$
and $F_{23}^{\beta}$ are different. But the term $\left(F_{23}^{\beta}-F_{23}^{\alpha}\right)\sigma_{32}$
vanishes due to $\sigma_{32}=0$, resulting in $W_{m}\equiv0$. 

Generalizing this example, it can be stated that the grains remain
thermodynamically identical during a variation of state of the system
when: 

(i) The differences $\left(U^{\beta}-U^{\alpha}\right)$, $\left(S^{\beta}-S^{\alpha}\right)$,
$\left(N_{k}^{\beta}-N_{k}^{\alpha}\right)$, $\left(n_{l}^{\beta}-n_{l}^{\alpha}\right)$
and $\left(V^{\beta}-V^{\alpha}\right)$ remain zero for all grain
regions with $N^{\beta}=N^{\alpha}$; 

(ii) The work $W_{m}$ of the transformation of one grain to the other,
given by Eq.~(\ref{eq:work1}), remains identically zero. \\
The term in $W_{m}$ with $i=3$ equals $\left(V^{\beta}-V^{\alpha}\right)\sigma_{33}$
and thus vanishes. As a result, the condition $W_{m}\equiv0$ reduces
to the identity 
\begin{equation}
{\displaystyle \sum_{i=1,2}}F_{11}F_{22}V^{\prime}\left(F_{i3}^{\beta}-F_{i3}^{\alpha}\right)\sigma_{3i}\equiv0.\label{eq:w0}
\end{equation}
As discussed in Section \ref{sub:phase-change}, the left-hand side
of this expression is the work of the shear stress along the transformation
vector $\mathbf{t}$ projected on the interface plane. This identity
is satisfied term by term when the components $F_{i3}$ are equal
($F_{i3}^{\beta}\equiv F_{i3}^{\alpha}$) for the directions $i$
in which the stress component $\sigma_{3i}$ is nonzero. In fact,
Eq.~(\ref{eq:w0}) reduces to this case after an appropriate rotation
of the coordinate axes. Note that coupled GBs\cite{Cahn06b} can be
equilibrated under stress as long as the relevant component of $\sigma_{3i}$
is zero. 

Under the above conditions, the phase-change equilibrium equation
(\ref{eq:deltaU_beta_alpha}) is satisfied as an identity. Furthermore,
it can be shown that when conditions (i) and (ii) are satisfied, Eqs.~(\ref{eq:for_gamma_bulk_1})
and (\ref{eq:for_gamma_bulk_b}) become identical to each other and
only one of them should be solved simultaneously with Eq.~(\ref{eq:gamma_global}).
As a result, $\gamma A$ is obtained by solving a system of only two
equations, giving

\begin{equation}
\begin{array}{ccl}
\gamma A & = & [U]_{X}-T[S]_{X}-{\displaystyle \sum_{k=2}^{K}}M_{k1}[N_{k}]_{X}-\phi_{1}[N]_{X}-{\displaystyle \sum_{l=1}^{L}}\mu_{l}[n_{l}]_{X}-{\displaystyle \sum_{i=1,2,3}}[V\overline{F}_{i3}/\overline{F}_{33}]_{X}\sigma_{3i},\end{array}\label{eq:gamma_GB}
\end{equation}
where

\begin{equation}
[Z]_{X}:=\frac{\left\vert \begin{array}{cc}
Z & X\\
Z^{\alpha} & X^{\alpha}
\end{array}\right\vert }{X^{\alpha}}=Z-Z^{\alpha}X/X^{\alpha}.\label{eq:Z_X_GB}
\end{equation}
Index $\alpha$ refers to one of the grains. By specifying $X$, one
term in Eq.~(\ref{eq:gamma_GB}) is eliminated. The coefficients
$[V\overline{F}_{i3}/\overline{F}_{33}]_{X}$ are the excess shears
($i=1$ or $2$) and excess volume $[V]_{X}$ ($i=3$) of the GB.

Similarly, if conditions (i) and (ii) are satisfied, then the Gibbs-Duhem
equations (\ref{eq: GD_global_a}) and (\ref{eq: GD_global_b}) for
the grains become identical to each other and the adsorption equation
is obtained by solving a system of only two equations:

\begin{equation}
\begin{array}{ccl}
d\left(\gamma A\right) & = & -[S]_{X}dT-{\displaystyle \sum_{k=2}^{K}}[N_{k}]_{X}dM_{k1}-[N]_{X}d\phi_{1}-{\displaystyle \sum_{l=1}^{L}[}n_{l}]_{X}d\mu_{l}\\
 & - & {\displaystyle \sum_{i=1,2,3}}[V\overline{F}_{i3}/\overline{F}_{33}]_{X}d\sigma_{3i}+{\displaystyle \sum_{i,j=1,2}}\tau_{ij}^{X}Ade_{ji}.
\end{array}\label{eq:The_AE_GB}
\end{equation}
Again, one variable in Eq.~(\ref{eq:The_AE_GB}) is eliminated by
specifying the extensive property $X$, which reduces the number of
independent differentials to $(K+L+6)$. The actual number of independent
variations is less due to the symmetry-related constraints imposed
for preservation of the identity of the grains. In the absence of
shear stresses, the last but one term in Eq.~(\ref{eq:The_AE_GB})
reduces to $[V]_{X}d\sigma_{33}$. In this particular case, Eq.~(\ref{eq:The_AE_GB})
can be applied to both coherent and incoherent symmetrical tilt boundaries.
We emphasize again that for Eqs.~(\ref{eq:gamma_GB}) and (\ref{eq:The_AE_GB})
to be valid, the condition $F_{i3}^{\beta}=F_{i3}^{\alpha}$ must
be satisfied for the directions $i$ in which $\sigma_{3i}\neq0$.

The last term in Eq.~(\ref{eq:The_AE_GB}) contains the GB stress
$\tau_{ij}^{X}$. Assuming that grain $\alpha$ is the reference state
of strain, it is straightforward to derive

\begin{eqnarray}
\tau_{ij}^{X} & = & \dfrac{1}{A}\left[V^{\prime}\overline{Q}_{ij}\right]_{X}\nonumber \\
 & = & \dfrac{1}{A}\left(\overline{\sigma}_{ij}V-\delta_{ij}\sigma_{33}V-AB_{i}\sigma_{3j}-\delta_{ij}{\displaystyle \sum_{k=1,2}}AB_{k}\sigma_{3k}\right)\label{eq:GB-stress}\\
 & - & \dfrac{X}{AX^{\alpha}}\left(\sigma_{ij}^{\alpha}V^{\alpha}-\delta_{ij}\sigma_{33}V^{\alpha}\right),\; i,j=1,2.\nonumber 
\end{eqnarray}
Here, $V$ is the bicrystal of volume, $\mathbf{B}$ is the displacement
vector of the upper boundary of the bicrystal during the GB formation
{[}cf. Fig.~\ref{fig:Creation-2}(d){]}, $\overline{\sigma}_{ij}$
is the volume-averaged stress tensor in the bicrystal, and all quantities
with superscript $\alpha$ refer an arbitrarily chosen homogeneous
region of grain $\alpha$. In the particular case when $X=N$ we have
\begin{eqnarray}
\tau_{ij}^{N} & =\dfrac{1}{A} & \left(\overline{\sigma}_{ij}V-\delta_{ij}\sigma_{33}V-AB_{i}\sigma_{3j}-\delta_{ij}{\displaystyle \sum_{k=1,2}}AB_{k}\sigma_{3k}\right)\nonumber \\
 & - & \dfrac{N}{AN^{\alpha}}\left(\sigma_{ij}^{\alpha}V^{\alpha}-\delta_{ij}\sigma_{33}V^{\alpha}\right),\; i,j=1,2.\label{eq:GB-stress-1}
\end{eqnarray}
For this choice of $X$, expression (\ref{eq:gamma_GB}) for $\gamma A$
takes the form
\begin{eqnarray}
\gamma A & = & [U]_{N}-T[S]_{N}-{\displaystyle \sum_{k=2}^{K}}M_{k1}[N_{k}]_{N}-{\displaystyle \sum_{l=1}^{L}}\mu_{l}[n_{l}]_{N}\nonumber \\
 & - & \sigma_{33}[V]_{N}-A{\displaystyle \sum_{i=1,2}}B_{i}\sigma_{3i}.\label{eq:gamma_GB-1}
\end{eqnarray}
Eqs.~(\ref{eq:GB-stress-1}) and (\ref{eq:gamma_GB-1}) will be used
in Part II of this work.\cite{Frolov2012a}

\section{Discussion and conclusions\label{sec:Discussion}}

We developed a thermodynamic theory of coherent solid-solid interfaces
in multicomponent systems under a general non-hydrostatic state of
stress. All equations were derived directly from the First and Second
Laws of thermodynamics. No Hooke's law or any other constitutive laws
of elastic deformation were invoked. No assumptions were made regarding
the interface structure other than the conservation of sites and elastic
response to applied shear stresses. 

To circumvent the problem of undefined chemical potentials of substitutional
components, we treat such components separately from interstitial
components using diffusion potentials introduced by Larchè and Cahn.\cite{Larche73,Larche_Cahn_78,Larche1985}
Diffusion potentials in non-hydrostatic solids are well-defined quantities
and, similar to chemical potentials, are uniform throughout an equilibrium
system. Because a system containing $K$ substitutional components
has only $(K-1)$ diffusion potentials {[}see Eq.~(\ref{eq:M}){]},
the requirement of their equality in coexisting phases ($M_{k1}^{\alpha}=M_{k1}^{\beta}$,
$k=2,...,K$) must be augmented by one more condition, namely, the
phase-change equilibrium equation discussed in Sections \ref{sub:Coherent-equilibrium}
and \ref{sub:phase-change}.

As an alternative to diffusion potentials, one can formulate the equilibrium
conditions in terms of the $\phi$-potentials introduced in this work
{[}Eq.~(\ref{eq:fi2_def}){]}. Equilibrium with respect to substitutional
components is then expressed by $K$ relations $\phi_{k}^{\alpha}=\phi_{k}^{\beta}$
($k=1,...,K$), which subsume the phase-change equilibrium condition.
Written in terms of the $\phi$-potentials, many equations of phase
equilibrium and interface thermodynamics look similar to the familiar
equations for fluid systems,\cite{Willard_Gibbs} with the $\phi$-potentials
playing the role of chemical potentials. In the particular case of
hydrostatically stressed solids, the $\phi$-potentials coincide with
real chemical potentials, which are well-defined thermodynamic quantities
under hydrostatic conditions. It should be emphasized, however, that
the $\phi$-potentials do not solve the problem of undefined chemical
potentials in non-hydrostatic solids in general. If we choose a different
interface plane between the same two phases, the $\phi$-potentials
will need to be redefined and will generally take on different numerical
values. 

An important result of this paper is the coherent phase coexistence
equation derived in Section \ref{sub:phase-coexist}. It can be written
as

\begin{equation}
\begin{array}{ccl}
0 & = & -\{S\}_{X}dT-{\displaystyle \sum_{k=1}^{K}}\{N_{k}\}_{X}d\phi_{k}-{\displaystyle \sum_{l=1}^{L}\{}n_{l}\}_{X}d\mu_{l}\\
 &  & -{\displaystyle \sum_{i=1,2,3}}\{VF_{i3}/F_{33}\}_{X}d\sigma_{3i}+{\displaystyle \sum_{i,j=1,2}}\{V^{\prime}Q_{ij}\}_{X}dF_{ji}
\end{array}\label{eq:bulk_coex-1}
\end{equation}
and is a generalization of the Clapeyron-Clausius equation to non-hydrostatically
stressed multicomponent systems. In the particular case when the phases
are hydrostatic, we have $Q_{ij}=0$, $\phi_{k}=\mu_{k}$, and this
equation reduces to 
\begin{equation}
0=-\{S\}_{X}dT-{\displaystyle \sum_{k=1}^{K+L}}\{N_{k}\}_{X}d\mu_{k}+\{V\}_{X}dp,\label{eq:hydro-coexit}
\end{equation}
where we use the same symbol $N_{k}$ for the amounts of substitutional
and interstitial components. For single-component phases this equation
recovers the classical Clapeyron-Clausius equation $0=-\{S\}_{N_{1}}dT+\{V\}_{N_{1}}dp$.\cite{Landau-Lifshitz-Stat-phys}
Eq.~(\ref{eq:bulk_coex-1}) predicts a rich variety of relations
between temperature, stress and chemical compositions of coexisting
phases, which call for testing by experiments and simulations in the
future.

Hydrostatic phase coexistence conditions and the Clapeyron-Clausius
equation have been extensively tested by experiment and simulations.
The conditions of coherent equilibrium between non-hydrostatic phases\cite{Robin1974,Larche73,Larche_Cahn_78}
(Section \ref{sub:Coherent-equilibrium}) and the phase coexistence
equation (\ref{eq:bulk_coex-1}) derived in this work call for a similar
verification. An analogue of Eq.~(\ref{eq:bulk_coex-1}) for solid-fluid
systems has been recently tested by atomistic simulations which implemented
strongly non-hydrostatic conditions in the solid.\cite{Frolov2010d,Frolov2010e}
A similar analysis and simulations could be performed for solid-solid
interfaces. In particular, it should be possible to test Eq.~(\ref{eq:bulk_coex_2})
by studying the relation between variations in the diffusion potential
and the shear stress.

We defined the interface free energy $\gamma A$ as the reversible
non-mechanical work performed in a thought experiment in which the
interface was formed by transforming a part of a single-phase system
into a different phase. We have shown that $\gamma A$ can be expressed
as excesses of different thermodynamic potentials, depending on the
choice of the reference properties $X$ and $Y$ in Eq.~(\ref{eq:gamma_excess_fi}).
Two examples of such potentials are given by Eqs.~(\ref{eq:gamma_excess_N})
and (\ref{eq:gamma_excess_N1}). 

Despite the freedom of choice in expressing $\gamma A$ as an excess,
it still remains the work of interface formation and therefore must
be unique. The following comment is due in this connection. The last
sum in Eqs.~(\ref{eq:gamma_excess}) and (\ref{eq:gamma_excess_fi})
is the mechanical work performed by the applied stress during the
phase transformation. This work depends on the transformation strain,
which we assumed to be known. However, crystal symmetry may lead to
multiplicity of possible transformation paths between the same two
phases. In other words, given the same reference state there can be
different deformation gradients $\mathbf{F^{\alpha}}$ and $\mathbf{F^{\beta}}$
producing exactly the same physical states of the phases. Some of
such alternate transformation paths can actually be implemented in
experiments or atomistic simulations. The situation is similar to
the existence of symmetry-dictated multiple modes of coupled GB motion.
Depending on the temperature and other factors, a moving GB can produce
different shear deformations of the receding grain, each corresponding
to a different mode of coupling. The existence of multiple coupling
modes was confirmed by both simulations\cite{Cahn06b,Cahn06a,Mishin06a,Suzuki05b}
and experiments\cite{Gorkaya2009,Molodov-2011a,Molodov-2011b,Molodov2011}
on symmetrical tilt grain boundaries.

The multiplicity of possible transformation strains seems to create
the following paradox: given a two-phase system in its current state,
how does one know which of the transformation strains should be plugged
in Eq.~(\ref{eq:gamma_excess_fi}) to obtain the correct value of
$\gamma A$? The answer lies in the fact that the phase equilibrium
equation (\ref{eq:bulk_coex}) also depends on the transformation
strain through the coefficients in front of $d\sigma_{3i}$. As discussed
in Section \ref{sub:phase-change}, one of the equilibrium conditions
is the neutrality with respect to spontaneous interface displacements
in an open system. The actual deformations of the phases occurring
during such displacements determine the equilibrium states of the
phases. Different deformation gradients will lead to different equilibrium
states. The equilibrium states of the phases, in turn, affect the
interface free energy and all other interface properties. If the actual
transformation strain realized in a given experiment or simulation
changes, so will the equilibrium states of the phases and thus $\gamma A$.
Thus, knowing the current state of the two-phase system, one should
be able to tell which of the symmetrically possible strains is actual
and thus correctly predict the interface properties.

Larchè and Cahn\cite{Larche73,Larche_Cahn_78,Larche1985} realized
that thermodynamic equilibrium between solid phases depends on the
degree of coherency of the interface. The above discussion suggests
that, for a coherent interface, the equilibrium also depends on the
particular coherent transformation strain (out of several possible
by crystal symmetry) which is realized during the interface displacements.
So do all thermodynamic properties of the interface. Mathematically,
the hypersurface of phase coexistence in the parameter space can have
multiple sheets corresponding to different transformation strains
between the same two phases. Switches between the transformation strains
can cause abrupt changes in states of the phases and thus in interface
properties, and are similar to interface phase transformations. The
strain multiplicity and its consequences for interface thermodynamics
are worth exploration in future work. One way to do so would be to
use Eq.~(\ref{eq:bulk_coex_2}) and apply the stress in different
directions which activate different transformation paths.

The generalized adsorption equation derived in this paper expresses
the differential of the interface free energy in terms of a set of
independent intensive parameters characterizing the equilibrium state
of a coherent two-phase system. Different forms of this equation are
given by Eqs.~(\ref{eq:the_AE}), (\ref{eq:the_AE-1}), (\ref{eq:the_AE_e}),
(\ref{eq:the_AE_e-1}), (\ref{eq:the_AE_phys}) and (\ref{eq: GH}).
The adsorption equation can be considered as the differential form
of the fundamental equation of the interface, whereas Eq.~(\ref{eq:gamma_excess_fi})
the fundamental equation is a functional form. The differential coefficients
in the adsorption equation define those interface excesses which are
measurable physical quantities. In particular, the interface stress
tensor $\boldsymbol{\mathbf{\mathbf{\mbox{\ensuremath{\tau}}}}}$
emerges from the coefficients in front of the lateral strains $e_{ij}$
and is formally defined by Eq.~(\ref{eq:tau_notat}). It should be
emphasized that the excess formulation of the interface stress presented
in this work is not as trivial as for hydrostatic systems. In the
latter case, the lateral stresses in the phases are equal and their
interface excess is unique. In non-hydrostatic systems, the lateral
stresses in the phases are different and their excess, if calculated
relative to a dividing surface, depends on its placement. In terms
of Cahn's generalized excesses,\cite{Cahn79a} $\boldsymbol{\mathbf{\mathbf{\mbox{\ensuremath{\tau}}}}}$
depends on the choice of the reference properties $X$ and $Y$. In
this sense, the interface stress between two non-hydrostatically stresses
solids in not unique. 

To understand the origin of this non-uniqueness, it is instructive
to consider its Lagrangian formulation (\ref{eq:stress-L}). As indicated
by Eq.~(\ref{eq:Cahn}), $\boldsymbol{\mathbf{\mathbf{\mbox{\ensuremath{\tau}}}}}_{L}$
is the response of the interface free energy $\gamma_{L}$ (per unit
reference area) to elastic deformations of the interface. Such deformations
must be implemented in such a way that to preserve the phase equilibrium.
In other words, the derivative in Eq.~(\ref{eq:Cahn}) must be taken
along a certain direction on the phase coexistence hypersurface in
the parameter space. Derivates taken along different directions give
generally different values of the interface stress, resulting in its
multiplicity. The direction in which we take the derivative is controlled
by the choice of the reference properties $X$ and $Y$.

Another excess quantity appearing in the generalized adsorption equation
is the interface excess shear. It characterizes the local elastic
shear deformation of the interface region in response to a shear stress
applied parallel to the interface. The excess shears $[V\overline{F}_{13}/\overline{F}_{33}]_{XY}$
and $[V\overline{F}_{23}/\overline{F}_{33}]_{XY}$ are conjugate to
the shear stress components $\sigma_{31}$ and $\sigma_{32}$, respectively.
Clearly, the excess shears are specific to coherent interfaces and
are undefined for interfaces that do not support shear stresses. This
explains why they did not appear in previous versions of the absorption
equation existing in the literature. For practical purposes, the excess
shears can be normalized by the physical or Lagrangian interface area,
giving the specific shears $[V\overline{F}_{i3}/\overline{F}_{33}]_{XY}/A$
and $[V\overline{F}_{i3}/\overline{F}_{33}]_{XY}/A^{\prime}$, respectively.
These quantities are similar to the GB ``slip'' introduced in the
context of the effective elastic response of GBs in polycrystalline
materials.\cite{Weissmueller2011} Excess shears of individual GBs
in copper calculated by atomistic methods will be reported in Part
II of this work.\cite{Frolov2012a}

The analysis presented in this work is limited to a plane interface
between semi-infinite homogeneous phases. In the future, some of these
constraints could be lifted by including, for example, the effect
of curvature or inhomogeneity. Such generalizations appear to be extremely
challenging but could benefit from the ideas and methods developed
previously in the mechanical theories of interfaces.\cite{Gurtin-1975,Gurtin1998,Fried1999}
As already mentioned, the mechanical theories consider only mechanical
equilibrium between the phases and do not impose the conditions of
chemical or phase equilibrium. They consider deformations of an \emph{already
existing} interface and disregard the process of its \emph{formation},
which, as we saw above and will see again in Part II,\cite{Frolov2012a}
can be accompanied by finite transformation strains. This prevents
the mechanical approach from properly defining $\gamma$ and deriving
the adsorption equation. Nevertheless, various interface excess stresses
and strains were identified, and carefully described mathematically,
under much more general conditions than in the present work. Besides
the excess shear and excess volume considered here, the mechanical
analysis reveals a few other excess strains arising, for example,
when the phases are curved and/or capable of relative rotation and/or
tilt.\cite{Gurtin1998} 

Finally, we note that our analysis neglects the existence of vacancies,
which is justified by their small concentration in most solids. In
the absence of vacancy sources and sinks in the system, the total
number of vacancies is conserved and they could be included in our
analysis as simply one of the substitutional components. In this case,
the number of vacancies could be treated as one of the independent
parameters alongside the amounts of real substitutional components.
As an alternate model, the vacancies can be assumed to be in equilibrium
with some sources and sinks existing far away from the interface.
In this case, the number of vacancies in any reference region of the
system is a \emph{dependent} parameter, whose value can be determined
from the condition of equilibrium with the sources and sinks. It should
be noted, however, that the vacancy equilibrium depends on specific
properties of the sources and sinks. For example, one can assume that
the phases terminate at surfaces parallel to the interface. Suppose
the surfaces are in contact with an inert atmosphere exerting a pressure
$p$ and are capable of absorbing and creating vacancies. The number
of vacancies in such a system is readily predictable and depends on
$p$ (Ref. \onlinecite{Herring1950}). However, the shear stresses
$\sigma_{31}$ and $\sigma_{32}$ have to be zero because of the presence
of surfaces. Thus, this model will not capture the interesting interface
properties associated with the shear stresses. 

On the other hand, a uniform distribution of vacancy sources and sinks
inside the phases, e.g. in the form of climbing dislocations, would
require a radical revision of the underlying assumptions of the present
analysis, particularly regarding the conservation of sites. This would
also raise the questions of possible creep deformation of the stressed
phases and the legitimacy of using the reference state formalism for
the description of elastic deformations. In view of these complications,
analysis of the possibility of incorporation of equilibrium vacancies
in thermodynamics of coherent interfaces is left for future work.
\begin{acknowledgments}
This work was supported by the U.S. Department of Energy, Office of
Basic Energy Sciences, the Physical Behavior of Materials Program,
under Grant No. DE-FG02-01ER45871.
\end{acknowledgments}

%\bibliographystyle{/Users/ymishin/YURI/Bibliography/aip}
%\bibliography{/Users/ymishin/YURI/Bibliography/literat}

\newpage{}\clearpage{}

\bigskip{}

\newpage{}\clearpage{}

\clearpage{}
\end{document}